\documentstyle[eqsecnum,floats,aps,prl,epsfig,twocolumn]{revtex}

\begin{document}

\wideabs{

\title { 
New theory of the $\gamma$-$\alpha$ phase transition in Ce:
quadrupolar ordering
} 
 
\author {A.V.~Nikolaev$^{1,2}$ and K.H.~Michel$^1$ \\} 

\address{
$^1$Department of Physics, University of Antwerp, UIA, 
2610 Antwerpen, Belgium\\
$^2$Institute of Physical Chemistry of RAS, 
Leninskii prospect 31, 117915, Moscow, Russia 
} 
 
\date{\today} 
 
\maketitle 
 
\begin{abstract} 
We present a theoretical model of the ``isostructural"
$\gamma-\alpha$ phase transition in Ce which is based on quadrupolar
interactions due to coupled charge density fluctuations
of 4$f$ electrons and of conduction electrons.
Conduction electrons are treated in tight-binding approximation.
The $\gamma-\alpha$ transition is described as an orientational
ordering of quadrupolar electronic densities in a $Pa{\bar 3}$
structure. The quadrupolar order of the conduction electron densities
is complementary to the quadrupolar order of 4$f$ electron
densities. The inclusion of conduction electrons
leads to an increase of the lattice contraction
at the $\gamma-\alpha$ transition in comparison to the
sole effect of 4$f$ electrons.
We calculate the Bragg scattering law and suggest synchrotron
radiation experiments in order
to check the $Pa{\bar 3}$ structure.
The theory is capable of accounting for transitions to phases
of non-cubic symmetry, but it is not sufficient to describe
the magnetic phenomena which we ascribe to the Kondo mechanism.
We also present a microscopic derivation of multipolar
interactions and discuss the crystal field of $\gamma$-Ce.
\end{abstract} 
 

}

\narrowtext

\section {Introduction} 
\label{sec:int} 

The understanding of the nature of the $\alpha$ phase and the
apparently isostructural transition between the cubic $\gamma$-
and $\alpha$- phases in cerium is a 
long-standing problem \cite{Kos,Mal}.
Numerous experimental data present an outstanding challenge
for the explanation by theoretical models 
\cite{Joh,Joh1,Sva,Jar,Laeg,All2,Lav,Gun,Bic}.
The most important question to be answered by theory is to find the
driving force of the ``isostructural" $\gamma-\alpha$ transition
and to explain also the existence of the other non-cubic phases of Ce.

The theory should also describe the change of
magnetic properties of Ce at the $\gamma-\alpha$ 
transition \cite{Mac,Kos} which are reminiscent of an
insulator-metal transition of 4$f$ electrons \cite{Joh}.
Recently the Mott transition has been reconsidered by
several authors using electronic band structure calculations 
with thermodynamic extensions \cite{Joh1,Sva,Jar}
(see for a review Ref. \cite{Laeg}).
There, the $\alpha$ phase is described as a regular band state formed
by 6$s$, 5$d$ and 4$f$ electronic states while in $\gamma$-Ce
different degrees of localization of 4$f$ states
are suggested and investigated. Two face centered cubic (fcc) 
phases of cerium
are attributed to two local minima of free energy
which develop for the same crystal structure ($Fm{\bar 3}m$).

Assuming the localized nature of the 4$f$ electrons throughout the
$\gamma-\alpha$ phase transition, one can understand 
the magnetic properties on the basis of two principles.
These are the singlet ground state of one 4$f$ electron and
the energy gap (characterized by the Kondo temperature $T_K$)
separating the ground state from a manifold of excited magnetic
states. So far these properties have been treated \cite{Gun,Bic}
on the basis of the Anderson
impurity Hamiltonian \cite{And} which implies an
antiferromagnetic Kondo spin interaction.
In order to describe the volume contraction \cite{Kos} at the
$\gamma-\alpha$ transition, the theoretical models \cite{All2,Lav,Laeg}
exploit the volume dependence of the Kondo temperature.
Such a ``volume collapse" leads to a phase instability without
symmetry change and is interpreted as an isostructural transition.
However within the Kondo theory scenario, the existence of
the other, non-cubic phases of Ce, remains unexplained.
In the last few years the validity of this approach has been
questioned by photoemission spectroscopy experiments 
\cite{Joy,Law,Bly,Andr,Ark} 
where the predicted temperature dependence of the intensity
of the Kondo resonance at the Fermi level has not
been observed. Recently it has appeared that the Kondo volume
collapse model can not be applied to YbInCu$_4$ \cite{Cor} which
exhibits a 0.5\% volume expansion during another isostructural phase
transition \cite{Sar} though Yb is the f-hole
analogue of Ce. In addition, taking into account the thermal
expansion of YbInCu$_4$ above the phase transition temperature $T=42$~K
one concludes that the Kondo temperature
is not a unique function of cell volume~\cite{Cor}.

An alternative theoretical model of the $\gamma-\alpha$
transition has been recently proposed by the present
authors \cite{NM}.
The theory is capable of accounting for transitions to
phases of non-cubic symmetry. 
The quantum mechanical electric quadrupole interaction between
4$f$ electrons on the fcc lattice
is proposed as the driving mechanism of a phase transition.
The $\gamma$-phase is characterized by the absence of spatial 
orientational order of the quadrupolar densities, the space 
group is $Fm{\bar 3}m$.
In the $\alpha$-phase the quadrupolar densities order in
a $Pa{\bar 3}$ structure. Notice that this change
from $Fm{\bar 3}m$ to $Pa{\bar 3}$ conserves
the fcc structure of the atomic center of mass points and
is solely due to orientational order of the quadrupoles.
This phase transition is accompanied by a contraction of
the fcc lattice, however the theoretical estimation of
these effects in \cite{NM} (in the following we will denote
this reference by I), is about an order of magnitude smaller
than the experimental result \cite{Kos}.
In addition the treatment of I does not indicate the
existence of a critical end point of the phase separation
line ($\gamma-\alpha$) in the $P-T$ (pressure-temperature)
phase diagram.

In the present paper, we will extend the theoretical model of I
by taking into account the polarization of $(6s5d)^3$
conduction band electrons.
The polarization can be considered as a screening process of 
the quadrupolar density orientations of the 4$f$ electrons
and results in turn in a complementary ordered $Pa{\bar 3}$ structure
build up from conduction electron quadrupolar densities.
The conduction electrons will be described within the
formalism of tight binding approximation.

The paper comprises the following sections.
We start (Sect.~2) with reconsidering 4$f$ electrons and extending
the treatment of I by taking into account the radial dependence
of the 4$f$ electron density.
Next in section~3 we describe the conduction electrons
in second quantization with basis functions in tight-binding
approximation. We derive the multipolar interactions
among conduction electrons as well as
interactions with 4$f$ electrons.
In considering conduction electrons we have to distinguish
between on- site and inter- site interactions.
Section~4 is devoted to a study of the crystal field
which acts on the individual 4$f$ electron. It is found that
the refinement by the radial density dependence of 4$f$
electrons does not improve but rather spoils the
agreement between experiment and theory. On the other hand
the inclusion of conduction electrons improves in turn
the situation.
In Sect.~5 we study the quadrupolar ordering of the coupled system
of 4$f$ electrons and conduction electrons.
We conclude that a state of lower free energy (in comparison
with the disordered state) can be achieved by a complementary
ordering of 4$f$ electron and conduction electron
quadrupolar densities.
On a same atomic lattice site, high charge density regions
of the 4$f$ electron correspond to low charge density
of the conduction electrons and vice versa.
Finally we present numerical estimates that
the inclusion of the conduction electron quadrupolar order
improves the magnitude of the lattice contraction
at the $Fm{\bar 3}m \rightarrow Pa{\bar 3}$ transition.
In order to propose an unambiguous experimental proof of
the present theoretical model, we calculate the Bragg
scattering law and suggest synchrotron radiation
experiments for the $Pa{\bar 3}$ structure (These calculations
will be presented in the full original version to be
published in Eur. Phys. J. B, 2000).
In Sect.~6 (Discussion) we recall the
salient features of the present theory and situate it with
respect to the conventional approaches that are based
on the Kondo theory concepts.
While we do not adopt the Kondo volume collapse models,
we discuss the relevance of the Friedel-Anderson hybridization
mechanism for the explanation of the magnetic anomalies
in Ce and suggest a link with our theory of the
electronic charge degrees of freedom driven structural
phase transition.

\section {Radial dependence of $4f$ electrons} 
\label{sec:rd} 

In our previous paper
\cite{NM} the quadrupolar coupling between 4$f$ electrons
has been calculated by assuming that the electron on
each lattice site $\vec{n}$ is localized on a sphere
with a fixed radius $r_f=1.378$ a.u.
Here we want to extend the previous calculation by
taking into account the radial dependence of the
4$f$ electron wave functions. Such an extension is
necessary if we want to study the interaction with
the conduction electrons (see next sections).
Furthermore it is useful in assessing 
the validity of our previous calculations.

We consider a face centered cubic crystal of $N$ Ce
atoms. Each atomic core possesses one 4$f$ electron.
In the $\gamma$-phase the 4$f$ electron densities are
orientationally disordered. The space group of the
crystal is $Fm{\bar 3}m$. 
The Coulomb interaction between two 4$f$ electrons
(charge $|e|$=1) at positions $\vec{R}(\vec{n})$
and $\vec{R}'(\vec{n}')$ near the lattice sites
$\vec{n}$ and $\vec{n}'$ reads
\begin{eqnarray}
  V(\vec{R}(\vec{n}),\vec{R}'(\vec{n}'))=
  {\frac{1}{|\vec{R}(\vec{n})-\vec{R}'(\vec{n}')|}}.
\label{2.1} 
\end{eqnarray}
The position vector $\vec{R}(\vec{n})$ is given by
\begin{eqnarray}
 \vec{R}(\vec{n}) = \vec{X}(\vec{n})+\vec{r}(\vec{n}).
\label{2.2} 
\end{eqnarray}
Here $\vec{X}(\vec{n})$ is the lattice vector which
specifies the centers of the atoms on a rigid fcc
lattice, while $\vec{r}(\vec{n})$ is the radius vector
of the 4$f$ electron; in spherical coordinates
$\vec{r}(\vec{n})=(r(\vec{n}),\Omega(\vec{n}))$,
where $\Omega=(\Theta,\phi)$. 
We perform a multipole expansion of
$V$ by using site symmetry adapted functions (SAF's)
\cite{Bra} which transform as irreducible representations
of the cubic site point group~$O_h$:
\begin{mathletters}
\begin{eqnarray}
  & &V(\vec{R}(\vec{n}),\vec{R}'(\vec{n}'))=
  \sum_{\Lambda \Lambda'} 
  v_{\Lambda \Lambda'}(\vec{n},\vec{n}';\,r,r')\,
  S_{\Lambda}(\vec{n})\, S_{\Lambda'}(\vec{n}'), \nonumber \\
 & & \label{2.3a} 
\end{eqnarray}
where
\begin{eqnarray}
  & & v_{\Lambda \Lambda'}(\vec{n},\vec{n}';\,r,r')\,
   =  \int \! d\Omega(\vec{n}) \int \! d\Omega(\vec{n}')\,
      {\frac{  S_{\Lambda}(\hat{n})\, S_{\Lambda'}(\hat{n}')}
     {|\vec{R}(\vec{n})-\vec{R}'(\vec{n}')|}} . \nonumber   \\
 & & \label{2.3b} 
\end{eqnarray}
\end{mathletters}
The SAF's $S_{\Lambda}(\hat{n})$, $\hat{n} \equiv \Omega(\vec{n})$ are
linear combinations of spherical harmonics $Y_l^m$ \cite{Bra}.
The index $\Lambda$ stands for $(l,\tau)$, with $\tau=(\Gamma,\mu,k)$.
Here $l$ accounts for the angular dependence of the
multipolar expansion, $\Gamma$ denotes an
irreducible representation (in the present case of the 
group $O_h$),
$\mu$ labels the representations that occur more than once and
$k$ denotes the rows of a given representation.
Expansion (\ref{2.3b}) still depends on the instantaneous radii
$r(\vec{n})$ and $r'(\vec{n}')$.
In Ref.~I we have written the Coulomb interaction in the
space of orientational state vectors $|i\rangle $, $i=1-14$, 
of the crystal field.
The wave functions $\langle\hat{n}|i\rangle $ 
were taken as linear combinations
of spin orbitals $Y_3^m(\Omega)u_s(s_z)$ $m=-3,...,+3$, where
$u_s$ is the spin function, with $s=\pm$ for the spin 
projections $s_z=\pm1/2$ on the $z$-axis respectively. 
(The consideration of spin orbitals is
necessary since we calculate the eigenvalues $\varepsilon_i$ of
the cubic crystal field in presence of spin-orbit coupling.)
Since at present we take into account the radial dependence of
the orbitals, we consider basis functions
\begin{eqnarray}
  \langle \vec{n},\vec{r}|i\rangle =
  {\cal R}_f(r(\vec{n}))\langle \hat{n}|i\rangle ,
\label{2.4} 
\end{eqnarray}
where we have assumed that the function ${\cal R}_f(r)$ 
is the same for all $i$.
The real radial function ${\cal R}_f(r)$ is obtained from
a DFT (density functional theory) calculation of an atom
of Ce within LDA (local density approximation) for $J=5/2$ states.
In Fig.~1 we plot the radial density for the outer
electrons.
%
\begin{figure} 
\centerline{
\epsfig{file=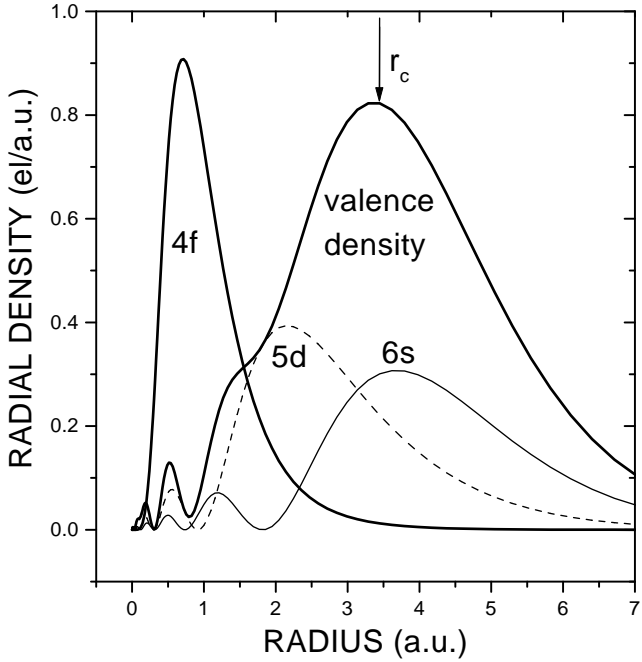,width=0.47\textwidth}
} 
\caption{
Calculated radial density distribution for 4$f$, 5$d$
and 6$s$ electrons of an isolated Ce atom.
Valence density is superposition
of $(6s)^2$ and 5$d$; $r_c$ is the close contact radius
of $\gamma$-Ce.
} 
\label{fig1} 
\end{figure} 
For a non-relativistic hydrogen-like atom ${\cal R}_f(r)$ 
would correspond to the
Laguerre function ${\cal R}_{n=4\,l=3}(r)$.

The matrix elements of the interaction (\ref{2.3a}) are 
obtained as
\begin{eqnarray}
  \langle i|_{\vec{n}}\langle i'|_{\vec{n}'}
  V(\vec{R}(\vec{n}),\vec{R}'(\vec{n}'))
  |j\rangle _{\vec{n}}|j'\rangle _{\vec{n}'} \nonumber \\
  =
  \sum_{\Lambda \Lambda'} 
  v_{\Lambda \Lambda'}^{F F}(\vec{n}-\vec{n}')\,
  c^F_{\Lambda}(ij)\, c^F_{\Lambda'}(i'j'), 
\label{2.5} 
\end{eqnarray}
where
\begin{eqnarray}
  v_{\Lambda \Lambda'}^{F F}(\vec{n}-\vec{n}') &=&
  \int \! dr\, r^2 \int \! dr'\, {r'}^2\,  \nonumber \\
  & &\times {\cal R}_f^2(r) {\cal R}_f^2(r')\,
  v_{\Lambda \Lambda'}(\vec{n},\vec{n}';\, r,r')
\label{2.6} 
\end{eqnarray}
accounts for the average radial dependence and where
\begin{eqnarray}
  c_{\Lambda}^F(ij) =  
  \int \! d\Omega \, \langle i|\hat{n}\rangle  
  S_{\Lambda}(\hat{n}) \langle \hat{n}|j\rangle .
\label{2.7} 
\end{eqnarray}
We use the superscript $F$ in order to indicate that
we have transitions between two 4$f$ states, {\it i.e.} $F \equiv (f,f)$.
By summing 
$V(\vec{R}(\vec{n}),\vec{R}'(\vec{n}'))$ over
all pairs of lattice sites $\vec{n}$, $\vec{n}'$, the total
Coulomb interaction operator is then obtained as
\begin{eqnarray}
  U^{ff}={\frac{1}{2}} {\sum_{\vec{n} \vec{n}'}}' \sum_{\Lambda \Lambda'}
  \rho_{\Lambda}^F(\vec{n}) \,
  v_{\Lambda \Lambda'}^{F F}(\vec{n}-\vec{n}') \,
  \rho_{\Lambda'}^F(\vec{n}'),
\label{2.8} 
\end{eqnarray}
where
\begin{eqnarray}
 \rho_{\Lambda}^F(\vec{n})=\sum_{ij} c^F_{\Lambda}(ij)
 |i\rangle _{\vec{n}}\langle j|_{\vec{n}}.
\label{2.9} 
\end{eqnarray}
Introducing Fourier transforms
\begin{mathletters}
\begin{eqnarray}
 & &\rho_{\Lambda}^F(\vec{q})= {\frac {1}{\sqrt{N}}}
 \sum_{\vec{n}} e^{i\vec{q} \cdot \vec{X}(\vec{n})}
 \rho_{\Lambda}^F(\vec{n}),  \label{2.10a} \\  
 & &v_{\Lambda \Lambda'}^{F F}(\vec{q})=  
 {\sum_{\vec{h} \neq 0}}' 
  e^{i\vec{q} \cdot \vec{X}(\vec{h})} 
  v_{\Lambda \Lambda'}^{F F}(\vec{h}), 
     \label{2.10b}  
\end{eqnarray}
\end{mathletters}
where $\vec{q}$ is the wave vector, we get
\begin{eqnarray}
 U^{ff}={\frac {1}{2}} \sum_{\vec{q}} \sum_{\Lambda \Lambda'}
 \rho_{\Lambda}^F(\vec{q})^{\dagger}  
 v_{\Lambda \Lambda'}^{F F}(\vec{q}) \, 
 \rho_{\Lambda'}^F(\vec{q}).
\label{2.11} 
\end{eqnarray}
The multipolar interaction (\ref{2.8}) or equivalently
(\ref{2.11}) can be separated into two parts.
We recall that $\Lambda \equiv (l,\tau)$.

Firstly we consider the case where $l \neq 0$ and $l'\neq0$. 
In I it has been shown that some of the coefficients
$c^F_{\Lambda}(ij) \equiv c_l^{\tau}(ij)$ for $l=2$ and
$\tau=(T_{2g},\mu=1,k=1-3)$ are different from zero.
(We recall that $i$ and $j$ refer to quantum states of the 4$f$ electron).
Therefrom we have inferred the existence of quadrupolar
($l=2$) density fluctuations caused by transitions among
4$f$ electron states.
In labeling the quadrupolar $T_{2g}$ functions, we recall
that the functions $S_2^k \equiv S_{(l=2,T_{2g},k)}$ are proportional to the
Cartesian components $yz$, $zx$ and $xy$ for $k=1$, 2 and 3
respectively.
In the basis of real spherical harmonics $Y_l^0$, $Y_l^{m,c}$
and $Y_l^{m,s}$ of Ref.~\cite{Bra} (see also Eqs.~(2.1a)-(2.1c) of I),
these functions correspond to $Y_2^{1,s}$, $Y_2^{1,c}$ and
$Y_2^{2,s}$. 
Writing only the index $k$ for $\Lambda=(l=2,k)$, we denote the
quadrupolar density operator by
\begin{eqnarray}
 \rho_k^F(\vec{n})=\sum_{ij} c^F_k(ij)
 |i\rangle _{\vec{n}}\langle j|_{\vec{n}}.
\label{2.12} 
\end{eqnarray}
The interaction between quadrupolar
4$f$ electron densities becomes
\begin{eqnarray}
 U_{QQ}^{ff}={\frac {1}{2}} \sum_{\vec{q}} \sum_{k k'}
 \rho_k^F(\vec{q})^{\dagger} 
 v_{k k'}^{F F}(\vec{q}) \, \rho_{k'}^F(\vec{q}).
\label{2.12a} 
\end{eqnarray}
The explicit form of $v_{k k'}^{F F}(\vec{q})$ is discussed in 
Appendix A.

Secondly we have the case where $l \neq 0$ and $l'=0$ or vice versa.
This means that we consider a multipole $l$ on a given lattice site
while the surrounding multipoles on neighboring fcc lattice sites
are taken in spherical approximation.
This interaction contributes to the crystal field.
The crystal field has unit cubic symmetry, the lowest nonzero
value of $l$ is 4 and $\tau=(A_{1g},1)$, where $A_{1g}$ is the
unit representation of the cubic site group $O_h$.
The crystal field contribution from 4$f$ electrons at site $\vec{n}$ 
is then given by
\begin{eqnarray}
  V_{CF}^f(\vec{n}) = \frac{12}{\sqrt{4\pi}}
  \sum_l v_{l, A_{1g}}^F {\;}_{0,A_{1g}}^F\; 
  \rho_{l, A_{1g}}^F \!(\vec{n}) .
\label{2.new14} 
\end{eqnarray}
Here we have restricted ourselves to the 12 nearest neighbors on
the fcc lattice, and $l=4,6,...$ .
 
The leading contributions to $U^{ff}$ are then given by
\begin{eqnarray}
  U^{ff}=U_{QQ}^{ff} + U_{CF}^{ff},
\label{2.new15} 
\end{eqnarray}
where $U_{CF}^{ff}=\sum_{\vec{n}} V_{CF}^f(\vec{n})$.
In the following of this section we will discuss the physical
consequences of the term $U_{QQ}^{ff}$, expression (\ref{2.12a}), 
also called orientational
pair quadrupolar interaction. We will give a discussion of
the crystal field in Sect.~4.

Previously (I) it has been found that the 
quadrupole-quadrupole interaction matrix 
$v^{F F}(\vec{q})$ 
becomes diagonal and has a twofold degenerate negative eigenvalue called
$\lambda_{X_5^+}$ at the $X$ point of Brillouin zone 
(BZ) of the 
fcc lattice (see also Appendix A).
This attractive interaction induces an orientational ordering
of the quadrupolar densities in a $Pa{\bar 3}$ structure.
A condensation scheme for the phase transition
$Fm{\bar 3}m \rightarrow Pa{\bar 3}$ reads:
\begin{mathletters}
\begin{eqnarray}
 & &{\bar \rho}_3^{F}(\vec{q}_x^X)={\bar \rho}_1^{F}(\vec{q}_y^X)
 ={\bar \rho}_2^{F}(\vec{q}_z^X)={\bar \rho}^F \sqrt{N} \neq 0, 
  \label{2.14a} \\
 & &{\bar \rho}_2^{F}(\vec{q}_x^X)={\bar \rho}_3^{F}(\vec{q}_y^X)
 ={\bar \rho}_1^{F}(\vec{q}_z^X)=0. 
  \label{2.14b} 
\end{eqnarray}
\end{mathletters}
Here the bar 
stands for a thermal expectation
value, while ${\bar \rho}^F$ is the order parameter amplitude.
The above condensation scheme corresponds to one of
eight possible domains of $Pa{\bar 3}$.
The wave vectors $\vec{q}_x^X$, $\vec{q}_y^X$ and $\vec{q}_z^X$
stand for $(2\pi/a)(1,0,0)$, $(2\pi/a)(0,1,0)$ and $(2\pi/a)(0,0,1)$
respectively, where $a$ is the cubic lattice constant.
For each arm of the star of 
${}^* \! \vec{q}^X=\{ \vec{q}_x^X, \vec{q}_y^X, \vec{q}_z^X \}$
there are two basis functions ${\bar \rho}_{k}^{F}$.
Hence the functions $\rho_k^F(\vec{q}^X)$ of the condensation
scheme (\ref{2.14a},b) form a basis of the six dimensional
irreducible representation $X_5^+$ of the space group
$Fm{\bar 3}m$. In real space the ordering implies four
sublattices of simple cubic structure as shown in figure 3
of Ref.~I. 

In I, where we have taken a fixed radius $r_f$
for the 4$f$ radial distribution, we found the eigenvalue
$\lambda_{X_5^+}=-3491$~K (Kelvin) and a phase transition
temperature $T_1=85.6$~K.
At present we have calculated the eigenvalue 
$\lambda_{X_5^+}=-4\gamma^{FF}$ (see Appendix A)
with the radial dependence ${\cal R}_f(r)$ in Eq.~(\ref{2.6})
taken from an atomic DFT calculation with LDA.
Although the atomic 4$f$ electron density is small beyond
the close contact radius $r_c=a/(2\sqrt{2})=3.448$~a.u., 
the opposite holds for the inter- site interaction
potential $v_{\Lambda \Lambda'}(\vec{n},\vec{n}';\, r,r')$
and its first derivative (see below), which increase
substantially when $r \rightarrow r_c$ and $r' \rightarrow r_c$
(see Fig.~2).
%
\begin{figure} 
\centerline{
\epsfig{file=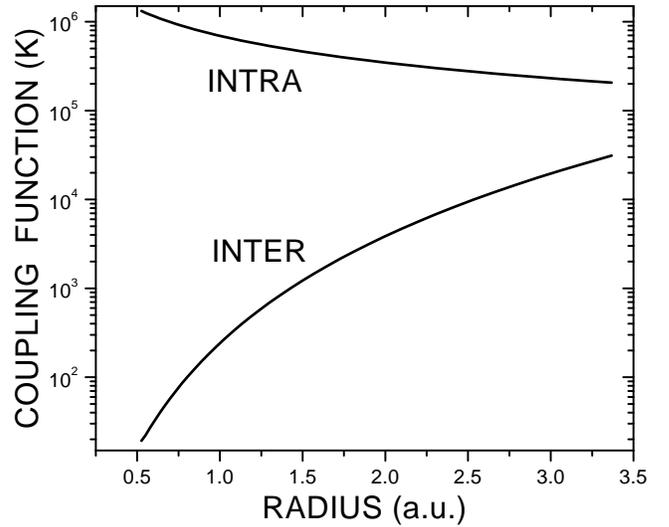,width=0.47\textwidth}
} 
\vspace{0.3cm}
\caption{
 Radial dependence of quadrupole-quadrupole 
 interactions~$v_{\Lambda \Lambda'}(\vec{n},\vec{n}';\,r,r')$,
 where $r=r'$, and 
 $\Lambda=\Lambda'=(l=2,T_{2g},k)$;
 $\vec{n}=\vec{n}'$ (intra-, $k=1,2,3$) and $\vec{n} \neq \vec{n}'$
 (inter- on fcc lattice) with $\vec{n}=(0,0,0)$,
 $\vec{n}'=(a/2)(0,1,1)$, $k=1$.
} 
\label{fig2} 
\end{figure} 
We have investigated several models for the radial
integral (\ref{2.6}).
In model 1 we consider the radial integration in the
range $0<r<r_c$, that is without any overlap of the
atomic 4$f$ electronic densities of neighboring atoms,
and we obtain $\lambda_{X_5^+}=-2121$~K,
next (model 2) we have allowed for an overlap between 
neighboring sites and extended the integration over the range
$0 \le r \le \infty$. We then obtain $\lambda_{X_5^+}=-2478$~K.
Finally (model 3) we assume again the integration range
$0 \le r \le r_c$ but renormalize the 4$f$ electronic density
to unity. The result for $\lambda_{X_5^+}$ is -2682~K.
Comparing these values with $\lambda_{X_5^+}=-3491$~K
obtained for the calculation with the characteristic
radius $r_f$, we conclude that a refinement of the theory
in smearing out the radial extension of the 4$f$ electron
distribution does not increase the strength of the
quadrupole-quadrupole interaction and consequently
does not increase the transition temperature. 

So far we have considered multipolar interactions on a rigid
lattice. In order to account for the lattice contraction
at the $\gamma-\alpha$ phase transition, we have to include
lattice displacements.
In I we have shown that the inter- site quadrupolar interaction
is modified by lattice displacements $u_{\nu}(\vec{n})$.
The correction to the potential reads
\begin{eqnarray}
  U_{QQT}&=&{\frac {1}{2}} {\sum_{\vec{n} \vec{n}'}}' \sum_{\nu}
  \sum_{k k'} v'_{\nu\, kk'}(\vec{n}-\vec{n}';r,r') \nonumber \\
  & &\times S_2^{k}(\hat{n}) \, S_2^{k'}(\hat{n}')\,
  \left[ u_{\nu}(\vec{n}) -u_{\nu}(\vec{n}') \right] .
\label{2.15} 
\end{eqnarray}
Here $S_2^{k}$ are SAF's with
$l=2$, $T_{2g}$, $k=1-3$. The coupling coefficients
$v'_{\nu\, kk'}$ are given by the derivative of the
quadrupole-quadrupole interaction with
respect to lattice displacements: 
\begin{eqnarray}
 v'_{\nu\, k k'}(\vec{n}-\vec{n}';\;r,r') 
  &=& \int \! d\Omega(\vec{n}) \int \! d\Omega(\vec{n}') \,
  S_2^{k}(\hat{n})\, S_2^{k'}(\hat{n}')   \nonumber \\
  & & \times 
  {\frac {\partial}{\partial X_{\nu}(\vec{n})}}
  {\frac{1}{|\vec{R}(\vec{n})-\vec{R}'(\vec{n}')|}}.
\label{2.16} 
\end{eqnarray}
Previously (I) this expression was considered for
4$f$ electrons on a shell with the (fixed) characteristic radius
$r=r'=r_f$.
Defining
\begin{eqnarray}
& & V'_{\nu}(\vec{R}(\vec{n}),\vec{R}'(\vec{n}'))=
 \sum_{k k'} v'_{\nu\, k k'}(\vec{n}-\vec{n}'; r,r') \,
 S_2^k(\hat{n}) \, S_2^{k'}(\hat{n}') \nonumber \\ 
& & \label{2.17} 
\end{eqnarray}
we now consider matrix elements with basis functions (\ref{2.4})
and obtain 
\begin{eqnarray}
  \langle i|_{\vec{n}}\langle i'|_{\vec{n}'}
  V'_{\nu}(\vec{R}(\vec{n}),\vec{R}'(\vec{n}'))
  |j\rangle _{\vec{n}}|j'\rangle _{\vec{n}'} \nonumber \\
  =\sum_{k k'} 
  v'_{\nu}{}_{k k'}^{F F}(\vec{n}-\vec{n}')\,
  c^F_k(ij) \, c^F_{k'}(i'j'),
\label{2.18} 
\end{eqnarray}
where
\begin{eqnarray}
 v'_{\nu}{}_{k k'}^{F F}(\vec{n}-\vec{n}') 
  &=& \int\! dr\, r^2 \int \! dr'\, {r'}^2\, 
  {\cal R}_f^2(r)\, {\cal R}_f^2(r')  \nonumber \\
  & &\times v'_{\nu\, k k'}(\vec{n}-\vec{n}';\;r,r').
\label{2.19} 
\end{eqnarray}
The correction to the quadrupolar interaction between
4$f$ electrons then becomes
\begin{eqnarray}
  U_{QQT}^{ff}={\frac {1}{2}} {\sum_{\vec{n} \vec{n}'}}'
  \rho_k^F(\vec{n}) \, v'_{\nu}{}_{k k'}^{F F}(\vec{n}-\vec{n}') \,
  \rho_{k'}^F(\vec{n}')       \nonumber \\
 \times\, \left[ u_{\nu}(\vec{n})-u_{\nu}(\vec{n}') \right], 
\label{2.20} 
\end{eqnarray}
with summation over repeated indices $k$, $k'$, $\nu$.
We introduce the Fourier expansion
\begin{eqnarray}
  u_{\nu}(\vec{n})=(Nm)^{-1/2}
  \sum_{\vec{q}} u_{\nu}(\vec{q})\,
  e^{i \vec{q} \cdot \vec{X}(\vec{n})} ,
\label{2.new23} 
\end{eqnarray}
where $m$ is the Ce mass.
Using definition (\ref{2.10a}),
we rewrite expression (\ref{2.20}) in Fourier space.
In the long wavelength limit $\vec{q} \rightarrow 0$
and taking $\vec{p}$ close to the star of $\vec{p}^X$,
we obtain
\begin{eqnarray}
 U_{QQT}^{ff}= i \sum_{\vec{p} \vec{q}} {\sum_{\nu(k)}}'
 v'_{\nu}{}_{k k}^{F F}(\vec{q},\vec{p})\,
 \rho_k^F(\vec{p})^{\dagger} \,
 \rho_k^F(\vec{p}) \, u_{\nu}(\vec{q}). 
 \label{2.22} 
\end{eqnarray}
Here the sum $\sum'$ refers to $\nu=x,y$ for $k$=3, to
$\nu=z,x$ for $k$=2 and to $\nu=z,y$ for $k$=1.
The coupling matrix is obtained as
\begin{eqnarray}
 & & v'_{\nu}{}_{3 3}^{F F}(\vec{q},\vec{p})=
 (Nm)^{-1/2} \Lambda^{F F}  q_{\nu} \, a
 \cos\left({\frac{p_x a}{2}}\right) 
 \cos\left({\frac{p_y a}{2}}\right), \nonumber \\ 
 & & \label{2.23} 
\end{eqnarray}
with $\Lambda^{F F}=v'_{\nu}{}_{3 3}^{F F}(\vec{n}-\vec{n}')$,
where $\nu=x$ or $y$ and 
$\vec{X}(\vec{n})-\vec{X}(\vec{n}')=(a/2)(1,1,0)$ on
the fcc lattice (for more details, see Ref.~I).
The other elements of $v'_{\nu}{}_{k k}^{F F}(\vec{q},\vec{p})$
follow by symmetry considerations and permutation of indices.

We consider expression  (\ref{2.22}) in the $Pa{\bar 3}$ ordered
phase by using the condensation scheme (\ref{2.14a}).
The lattice displacements are taken in the long wavelength limit
where they are related to the homogeneous strains.
Symmetry implies that only longitudinal strains occur:
\begin{eqnarray}
   \lim_{\vec{q} \rightarrow 0} iq_{\nu}{\bar u}_{\nu}(\vec{q})
   =\sqrt{mN} \epsilon_{\nu \nu},\;\;\; \nu=x,y,z.
\label{2.new26}
\end{eqnarray}
Then $U_{QQT}$ becomes
\begin{eqnarray}
  \frac{1}{N} U_{QQT}^{ff}=-2a \Lambda^{F F} \, ({\bar \rho}^F)^2
  \sum_{\nu} \epsilon_{\nu \nu} , 
\label{2.new27}
\end{eqnarray}
which corresponds to a coupling of ordered quadrupoles
(quadratic) to longitudinal strains.
Notice that the sign of $\Lambda^{F F}$ is negative,
this is a consequence of the repulsive nature of
quadrupole-quadrupole interaction.
The strains give rise to an elastic energy of the
cubic lattice:
\begin{eqnarray}
  \frac{1}{N} U_{TT}& =& \frac{a^3}{4} \left[
  c_{11}^0(\epsilon_{xx}^2+\epsilon_{yy}^2+\epsilon_{zz}^2)
  \right. \nonumber \\
  & & \left. +2c_{12}^0(\epsilon_{xx} \epsilon_{yy} +
  \epsilon_{yy} \epsilon_{zz} + \epsilon_{zz} \epsilon_{xx}) \right] .
\label{2.new28}
\end{eqnarray}
Here $c_{11}^0$ and $c_{12}^0$ are the bare elastic constants.
The interplay of quadrupolar order and lattice displacements
follows from the interaction Hamiltonian
\begin{eqnarray}
 U=U_{TT}+U_{QQ}^{ff}+U_{QQT}^{ff}.
 \label{2.new29} 
\end{eqnarray}
Minimizing $U[{\bar \rho}^F,\epsilon_{\nu \nu}]$ with respect to the strains
for a given $Pa{\bar 3}$ ordered structure, we obtain
$\epsilon_{\nu \nu}$, 
\begin{eqnarray}
 \epsilon_{xx}=\epsilon_{yy}=\epsilon_{zz}=
 -8a^{-2}\, | \Lambda^{F F} | \, \kappa_L \, ({\bar \rho}^F)^2 ,
 \label{2.new30} 
\end{eqnarray}
while the change of the lattice constant is given by 
$\triangle a=\epsilon_{xx} a$.
Here $\kappa_L=(c_{11}^0+2c_{12}^0)^{-1}$ is the linear
compressibility.
Hence the present theory leads unambiguously to a lattice
contraction. In order to provide a numerical estimate of the
lattice contraction, one has to calculate $\Lambda^{F F}$,
to estimate $\kappa_L$ from experimental results and to
calculate by the methods of statistical mechanics 
the discontinuity of the order parameter at the $\gamma-\alpha$
phase transition. In I (fixed radius $r_f$) we did obtain 
$\Lambda^{F F}=-445$~K/a.u. ($-$841 K/{\AA}) \cite{err1}. Now,
for the models 1), 2), and 3) of the spatial
radial integrals we obtain $\Lambda^{F F}$=$-498$, $-581$ and
$-659$~K/{\AA} respectively.
We conclude that the present refinement in calculating
$\Lambda^{F F}$ is not helpful in view of obtaining a larger
value of $\Lambda^{F F}$ and hence of the lattice contraction
$\triangle a$.
Since our theoretical value of $\triangle a$ is more
than one order of magnitude too small to account for
the $\sim 15$\% volume contraction of cerium, we
conclude that a significant process has so far been
omitted in our treatment.
In the following section we will study the effect of
conduction electrons on quadrupolar interactions. 

\section {Multipolar interactions} 
\label{sec:tb} 

Here we will investigate about the existence of multipolar
interactions between the localized 4$f$ electrons and the
$(6s5d)^3$ conduction electrons $U^{fc}$ as well as multipolar
interactions among conduction electrons $U^{cc}$.

As a consequence of Bloch's theorem  the conduction
electron states are classified according to the irreducible
representations of the translational symmetry group of
the crystal. Nevertheless, in the proximity of the nuclei
the corresponding wave functions can be expanded
in terms of spherical harmonics. This fact reflects the
importance of Coulomb singularities associated with the
nuclei. In the following we will focus on the interactions
inside the ``muffin-tin" or touching spheres centered on
the nuclei and for the description of itinerant states
adopt the tight-binding approximation.
In absence of a static magnetic field, conduction electronic
states with spin projections $s_z=\pm 1/2$ are degenerate.
Hence we will omit the spin dependence of the wave
function.
The wave function of a conduction electron with wave vector
$\vec{k}$ and band index $\alpha$ is then written as a linear
combination of local atomic wave functions 
\begin{eqnarray}
& & \langle \vec{R} | \, \vec{k},\alpha \rangle =
 \psi_{\vec{k},\alpha}(\vec{R}) \nonumber \\
& & =\frac{1}{\sqrt{N}}
 \sum_{\vec{n}'} e^{i\vec{k}\vec{X}(\vec{n}')}
 \sum_{lm} \gamma_{lm}(\vec{k},\alpha) \,
 \phi_{lm}(\vec{R}-\vec{X}(\vec{n}')),
\label{3.1} 
\end{eqnarray}
where the position vector $\vec{R}$ is given by Eq.~(\ref{2.2}).
The atomic wave functions are given by 
$\phi_{lm}(\vec{r})={\cal R}_l(r) Y_l^m(\Omega)$.
The expansion coefficients $\gamma_{lm}(\vec{k},\alpha)$ and
corresponding eigenvalues $E(\vec{k},\alpha)$ are obtained by
solving the secular equation (see e.g. \cite{Ash})
\begin{eqnarray}
 \sum_{\lambda} \left({H}_{\lambda\,\lambda'}(\vec{k})-
 E(\vec{k},\alpha)\, {S}_{\lambda\,\lambda'}(\vec{k}) \right) 
 \gamma_{\lambda'}(\vec{k},\alpha) = 0.
\label{3.2} 
\end{eqnarray}
Here ${H}(\vec{k})$ and ${S}(\vec{k})$ are
matrices of single particle Hamiltonian and overlap,
respectively.
Here and in the following $\lambda=(l,m)$, 
$\delta_{\lambda \lambda'}=\delta_{l l'}\delta_{m m'}$.
The eigenvalues $E(\vec{k},\alpha)$ refer to the energy band spectrum
of the conduction electrons.
Notice that the wave function $\psi_{\vec{k},\alpha}(\vec{R})$
satisfies the Bloch condition. In the following we shall use
an idealized basis set of orthogonal Wannier functions
$\phi_{\lambda}(\vec{r}(\vec{n}))=\langle \vec{R}| \lambda\rangle _{\vec{n}}$
without overlap, 
\begin{eqnarray}
 {}_{\vec{n}} \langle \lambda|\lambda'\rangle _{\vec{n}'}=
 \delta_{\lambda \lambda'} \delta_{\vec{n} \vec{n}'}.
\label{3.3} 
\end{eqnarray}
The coefficients $\gamma_{\lambda}$ satisfy the relation
\begin{eqnarray}
  \sum_{\lambda} \gamma_{\lambda}^*(\vec{k},\alpha) 
  \gamma_{\lambda}(\vec{k}',\alpha')=
  \delta_{\vec{k} \vec{k}'} \delta_{\alpha \alpha'}.
\label{3.4} 
\end{eqnarray}
The functions $\psi_{\vec{k},\alpha}(\vec{R})$ are then normalized:
\begin{eqnarray}
  \langle \vec{k},\alpha|\vec{k}',\alpha'\rangle =
  \delta_{\vec{k} \vec{k}'} \delta_{\alpha \alpha'}.
\label{3.5} 
\end{eqnarray}
The corresponding electronic Hamiltonian in second quantization is
\begin{eqnarray}
 U_0^c=\sum_{\vec{k} \alpha} E_{\vec{k} \alpha} 
  a^{\dagger}_{\vec{k} \alpha} a_{\vec{k} \alpha} ,  
\label{A.6} 
\end{eqnarray}
where
$a_{\vec{k} \alpha}^{\dagger}$ and $a_{\vec{k} \alpha}$
are creation and annihilation operators for one electron
in state $(\vec{k},\alpha)$.

In case of cerium, we will restrict ourselves to conduction electrons
of $d$ type ($l=2$, $m$=-2,...,+2, with principal quantum number
$n$=5) and of $s$ type ($l$=0, $m$=0, with $n$=6).
Fig.~1 shows the radial density of the valence electrons
of a Ce atom, it should also be a guide
of the relative spatial extension of the localized 4$f$ and
the conduction electrons in the crystal.

Although the derivation of multipolar interactions presented later 
in this Section
is general in the following we imply the high temperature 
$\gamma$ phase of cerium.
There the total electronic density associated with the conduction
electrons has the symmetry of the space group $Fm{\bar 3}m$ \cite{NM}.
At each lattice site the band electronic density has on average the
unit symmetry ($A_{1g}$) of the cubic point group $O_h$.
It is important to realize that the on-site quadrupolar
charge density fluctuations of $T_{2g}$ and $E_g$ symmetry
do not interact with the totally symmetric charge distribution
($A_{1g}$) representing the ground state of cerium
(see also Eq.~(\ref{3.10})).
On one hand this allows us to use the Landau concept of Fermi
liquid and to consider the conduction electrons described by
Eq.~(\ref{3.1}) as quasiparticles where the ground state on-site
interactions have already been taken into account through
the coefficients $\gamma_{lm}(\vec{k},\alpha)$.
On the other hand we can focus only on the quadrupolar interactions
where conduction electrons are involved and consider these
interactions separately from the ground state.

Interactions involving conduction electrons will be described
within the formalism of second quantization.
We introduce field operators
\begin{mathletters}
\begin{eqnarray}
  & &\Psi(\vec{R})=\sum_{\vec{k} \alpha} a_{\vec{k} \alpha}
   \langle \vec{R}| \vec{k} \alpha\rangle ,  \label{3.6a} \\
  & &\Psi^{\dagger}(\vec{R})=\sum_{\vec{k} \alpha} 
   a_{\vec{k} \alpha}^{\dagger}
   \langle \vec{k} \alpha |\vec{R}\rangle .  \label{3.6b}  
\end{eqnarray}
\end{mathletters}
The operators $\Psi$ and $\Psi^{\dagger}$ satisfy the usual
anticommutation relations for fermion fields. 
Although the derivation of multipolar expansions for these
interactions is analogous, in particular as far as symmetry
is concerned, to the procedure of Sect.~2, there is an
essential difference. Since the conduction electrons are not localized, 
they give rise to quadrupolar pair interactions on a same site
(intra) and to interactions between different sites (inter).

The Coulomb interaction between localized 4$f$ electrons
at sites $\{ \vec{n} \}$ and the conduction electrons is
given by
\begin{eqnarray}
   U^{fc}&=&\sum_{\vec{n}} \sum_{ij} |i\rangle _{\vec{n}}\langle j|_{\vec{n}}
   \nonumber \\
   & &\times \int \! d\vec{R}'\, \Psi^{\dagger}(\vec{R}')
   \langle i|_{\vec{n}} V(\vec{R}(\vec{n}),\vec{R}')|j\rangle _{\vec{n}}
   \Psi(\vec{R}').
\label{3.7} 
\end{eqnarray}
We observe that 
$\int d \vec{R}' \rightarrow \sum_{\vec{n}'} \int d \vec{r}{\,}'$,
where the integral extends over the volume of the cell $\vec{n}'$.
In expression (\ref{3.7}) we perform a multipole expansion of $V$
similar to Eqs. (\ref{2.3a},b) and then calculate matrix elements
\begin{eqnarray}
 \langle i|_{\vec{n}} \langle \vec{k},\alpha|  
  V(\vec{R}(\vec{n}),\vec{R}'(\vec{n}'))
  |j\rangle _{\vec{n}}  
  |\vec{p},\beta\rangle   =
  {\frac {e^{i(\vec{p}-\vec{k}) \cdot \vec{X}(\vec{n}')}}{N}}
  \nonumber \\
  \times\, \sum_{\Lambda \Lambda'} \sum_{l_1 l_2} 
  v_{\Lambda}^{F}{\,}_{\Lambda'}^{l_1 l_2}(\vec{n}-\vec{n}')\,
  c^F_{\Lambda}(ij) \,
  c_{\Lambda' l_1 l_2}(\vec{k},\alpha;\,\vec{p},\beta) .
  \nonumber \\
\label{3.8} 
\end{eqnarray}
Here we define
\begin{mathletters}
\begin{eqnarray}
  v_{\Lambda}^{F}{\,}_{\Lambda'}^{l_1 l_2}(\vec{n}-\vec{n}')& &=
  \int\! dr r^2 \int \! dr' {r'}^2 \,
  {\cal R}_f^2(r) \nonumber \\
  & & \times v_{\Lambda \Lambda'}(\vec{n},\vec{n}';r,r') \,
  {\cal R}_{l_1}(r')\, {\cal R}_{l_2}(r'),
   \label{3.9a} 
\end{eqnarray}
\begin{eqnarray}
  c_{\Lambda'\, l_1 l_2}(\vec{k},\alpha;\,\vec{p},\beta)&=&
  \sum_{m_1\,m_2}
  \gamma_{l_1 m_1}^*(\vec{k},\alpha) \, \gamma_{l_2 m_2}(\vec{p},\beta) 
  \nonumber \\
  & & \times \, c_{\Lambda'}(l_1 m_1,\,l_2 m_2),
   \label{3.9b}
\end{eqnarray}
and
\begin{eqnarray}
 & & c_{\Lambda'}(l_1 m_1,\,l_2 m_2)=\int Y_{l_1}^{m_1\,*}(\Omega)\, 
 S_{\Lambda'}(\Omega) \, Y_{l_2}^{m_2}(\Omega)\, d\Omega. \nonumber \\
 & & \label{3.9c}
\end{eqnarray}
\end{mathletters}
In the following we will introduce a single index $L$ for $(l_1,l_2)$,
writing $v_{\Lambda\, \Lambda'}^{F\, L}$ for 
$v_{\Lambda}^{F}{\,}_{\Lambda'}^{l_1 l_2}$, $c_{\Lambda'\, L}$
for $c_{\Lambda'\, l_1 l_2}$ etc..
We notice that the index $l_1(l_2)$ takes the values 0 and 2
corresponding to $s$ and $d$ electrons.
We recall that $c^F_{\Lambda}(ij)$, referring to the 4$f$ electron
transitions, is given by expression (\ref{2.7}).

Since the conduction electrons are delocalized, we will have
to distinguish between interactions where $\vec{n} \neq \vec{n}'$
(inter-site), and where $\vec{n} = \vec{n}'$ (on-site).
In the first case $v_{\Lambda \Lambda'}(\vec{n},\vec{n}';\,r,r')$
is given by an expression of type (\ref{2.3b}), in the
second case we have
\begin{eqnarray}
  v_{\Lambda \Lambda'}(\vec{n}=\vec{n}';\,r,r')\,
  &=& \int \! d\Omega  \int \! d\Omega'  
  {\frac{1}{|\vec{r} -\vec{r}{\,}' |}}
  S_{\Lambda}(\Omega) \, S_{\Lambda'}(\Omega') \nonumber \\
  &=&\left( {\frac {r^l_< }{r^{(l+1)}_> }} \right)
  {\frac {4\pi}{2l+1}} 
  \delta_{\Lambda \Lambda'} ,
\label{3.10} 
\end{eqnarray}
which is independent of the site, as is also the case
for $v_{\Lambda}^{F}{\,}_{\Lambda'}^{L}(\vec{n}=\vec{n}')$.
Here $r_>=max(r,r')$, $r_<=min(r,r')$ and 
$\delta_{\Lambda \Lambda'}=\delta_{\tau \tau'} \delta_{l l'}$.
The inter-site coupling
$v_{\Lambda}^{F}{\,}_{\Lambda'}^{L}(\vec{n}-\vec{n}')$ 
still depends on the distance
$|\vec{X}(\vec{n})-\vec{X}(\vec{n}')|$, as follows from the
translational invariance of the lattice.

In addition to the multipole density of 4$f$ electrons
$\rho_{\Lambda}^F(\vec{q})$, Eq.~(\ref{2.10a}), we define
the multipole density of conduction electrons 
\begin{mathletters}
\begin{eqnarray}
 & &\rho_{\Lambda}^{L}(\vec{n}) =
 {\frac {1}{\sqrt{N}}} \sum_{\vec{q}}
 \rho_{\Lambda}^{L}(\vec{q}) \, e^{-i\vec{q} \cdot \vec{X}(\vec{n})},
\label{3.11a} \\ 
 & &\rho_{\Lambda}^{L}(\vec{q}) =
 {\frac {1}{\sqrt{N}}} \sum_{\alpha \beta} \sum_{\vec{k}}
 a_{\vec{k} \alpha}^{\dagger} a_{\vec{k}-\vec{q} \beta}\,
 c_{\Lambda\, L} (\vec{k},\alpha; \vec{k}-\vec{q},\beta).
 \nonumber \\
\label{3.11b} 
\end{eqnarray}
\end{mathletters}
The interaction for the inter- site contribution of $U^{fc}$
is then given by
\begin{mathletters}
\begin{eqnarray}
 \left. U^{fc} \right|_{inter}= 
 \sum_{\vec{q}}\, \rho_{\Lambda}^F(\vec{q})^{\dagger}\, 
 v_{\Lambda}^{F}{\,}_{\Lambda'}^{L}(\vec{q})\,
 \rho_{\Lambda'}^{L}(\vec{q}),
\label{3.12a} 
\end{eqnarray}
with
\begin{eqnarray}
  v_{\Lambda}^{F}{\,}_{\Lambda'}^{L}(\vec{q})=
  {\sum_{\vec{h} \neq 0}}' 
  e^{i \vec{q} \cdot \vec{X}(\vec{h}) } \,
  v_{\Lambda}^{F}{\,}_{\Lambda'}^{L}(\vec{h}). 
\label{3.12b} 
\end{eqnarray}
\end{mathletters}
The on-site part of $U^{fc}$ is obtained as
\begin{mathletters}
\begin{eqnarray}
 \left. U^{fc} \right|_{intra}=
 C_{\Lambda}^{F}{\,}_{\Lambda}^{L}\,
 \sum_{\vec{q}}\, \rho_{\Lambda}^F(\vec{q})^{\dagger} \, 
 \rho_{\Lambda}^{L}(\vec{q}),
\label{3.12'a} 
\end{eqnarray}
where
\begin{eqnarray}
  C_{\Lambda}^{F}{\,}_{\Lambda}^{L}=
  v_{\Lambda}^{F}{\,}_{\Lambda}^{L}(\vec{n}=\vec{n}').
\label{3.12'b} 
\end{eqnarray}
\end{mathletters}
Here we have also used the orthogonality relation (\ref{3.10}).
In expressions (\ref{3.12a}) and (\ref{3.12'a}), summation is
understood over indices $\Lambda$, $\Lambda'$, $L$ ($l_1$, $l_2$).

In a similar way we treat the multipolar interactions
between conduction electrons.
We now start from the expression in operator representation 
\begin{eqnarray}
  & & U^{cc}={\frac {1}{2}} \int \! d \vec{R}' \! \int \! d \vec{R} \,
  \Psi^{\dagger}(\vec{R}') \Psi^{\dagger}(\vec{R})\,
  V(\vec{R}, \vec{R}')\,
  \Psi(\vec{R}) \Psi(\vec{R}') . \nonumber \\
 & & \label{3.14} 
\end{eqnarray}
We have to consider matrix elements
\begin{eqnarray}
 & & \langle \vec{k},\alpha| \langle \vec{k}',\alpha'| 
  V(\vec{R},\vec{R}')
  |\vec{p},\beta\rangle |\vec{p'},\beta'\rangle = \nonumber \\
 & & \sum_{\vec{n} \vec{n}'} \sum_{\Lambda \Lambda'} 
  \sum_{L} \sum_{L'}
  v_{\Lambda}^{L}{\;}_{\Lambda'}^{L'}(\vec{n}-\vec{n}')\,
  c_{\Lambda L}(\vec{k},\alpha;\,\vec{p},\beta)
  \nonumber \\
  & & \times  
  c_{\Lambda' L'}(\vec{k}',\alpha';\,\vec{p'},\beta')\,
  {\frac {e^{i(\vec{p}-\vec{k})\vec{X}(\vec{n})}}{N}}
  {\frac {e^{i(\vec{p'}-\vec{k}')\vec{X}(\vec{n}')}}{N}},
\label{3.15} 
\end{eqnarray}
where
\begin{eqnarray}
  v_{\Lambda}^{L}{\,}_{\Lambda'}^{L'}
  (\vec{n}&-&\vec{n}')=
  \int dr\, r^2 \int dr'\, {r'}^2\,
   {\cal R}_{l_1}(r) {\cal R}_{l_2}(r) \nonumber \\
  & & \times \, v_{\Lambda \Lambda'}(\vec{n},\vec{n}';r,r')\,
  {\cal R}_{l'_1}(r') {\cal R}_{l'_2}(r').
   \label{3.16} 
\end{eqnarray}
Again we distinguish inter- site and intra- site interactions.
We obtain for the inter- site contribution
\begin{mathletters}
\begin{eqnarray}
 \left. U^{cc} \right|_{inter} &=& {\frac {1}{2N}}
 \sum_{\vec{k} \vec{k}'\, \vec{q}}
 a_{\vec{k} \alpha}^{\dagger} a_{\vec{k}' \alpha'}^{\dagger}
 a_{\vec{k}'-\vec{q} \beta} a_{\vec{k}+\vec{q} \beta'}\, 
    \nonumber  \\
 &\times & 
 v_{\Lambda}^{L}{\,}_{\Lambda'}^{L'}(\vec{q})\,
 c_{\Lambda\, L}(\vec{k}, \alpha; \vec{k}+\vec{q}, \beta)
 \label{3.17a} \\ 
 & \times &
 c_{\Lambda'\, L'}(\vec{k}', \alpha' ; \vec{k}'-\vec{q}, \beta'),
 \nonumber 
\end{eqnarray}
with
\begin{eqnarray}
  v_{\Lambda}^{L}{\,}_{\Lambda'}^{L'}(\vec{q})=
  {\sum_{\vec{h} \neq 0}}' 
  e^{i \vec{q} \cdot \vec{X}(\vec{h})}\, 
  v_{\Lambda}^{L}{\,}_{\Lambda'}^{L'}(\vec{h}).
\label{3.17b} 
\end{eqnarray}
\end{mathletters}
In Eq.~(\ref{3.17a}) summation is understood over the indices
$\alpha...$, $L$($l_1,l_2$)... , $\Lambda...$ .
Expression (\ref{3.17a}) can be rewritten as
\begin{eqnarray}
 \left. U^{cc} \right|_{inter}={\frac {1}{2}}
 \sum_{\vec{q}} \,  
 v_{\Lambda}^{L}{\,}_{\Lambda'}^{L'}(\vec{q})\;
 \eta \! \left(
 \rho_{\Lambda}^{L}(\vec{q})^{\dagger} 
 \rho_{\Lambda'}^{L'}(\vec{q})
 \right), \nonumber \\
\label{3.18} 
\end{eqnarray}
where $\rho_{\Lambda}^{L}(\vec{q})$ is given by
Eq.~(\ref{3.11b}) and where $\eta$ is the normally
ordered product operator such that all the $a^{\dagger}$
are placed to the left and all $a$ to the right in the product.

The on-site contribution is given by
\begin{mathletters}
\begin{eqnarray}
 \left. U^{cc} \right|_{intra}= {\frac {1}{2}}
 C_{\Lambda}^{L}{\,}_{\Lambda}^{L'}\,
 \sum_{\vec{q}}\,  
 \eta \! \left(
 \rho_{\Lambda}^{L}(\vec{q})^{\dagger} 
 \rho_{\Lambda}^{L'}(\vec{q})
 \right), \nonumber \\ 
\label{3.19a}  
\end{eqnarray}
with
\begin{eqnarray}
  C_{\Lambda}^{L}{\,}_{\Lambda}^{L'}=
  v_{\Lambda}^{L}{\,}_{\Lambda}^{L'}(\vec{n}=\vec{n}').
\label{3.19b} 
\end{eqnarray}
\end{mathletters}
So far the present formalism is general as far as multipoles are
concerned.  
In the following of this section we will study the interaction
between quadrupoles ($l=2$, $l'=2$). In Sect.~4 we will study
the crystal field ($l=4$, $l'=0$).

We consider the three quadrupolar components of $T_{2g}$
symmetry and write the index $k$ for $\Lambda\,=\,(l=2,k)$,
with $k=1-3$.
The quadrupolar density of conduction electrons becomes
\begin{eqnarray}
 \rho_{k}^{L}(\vec{q}) =
 {\frac {1}{\sqrt{N}}} \sum_{\alpha \beta} \sum_{\vec{p}}
 a_{\vec{p} \alpha}^{\dagger} a_{\vec{p}-\vec{q} \beta}\,
 c_{k\, L} (\vec{p},\alpha; \vec{p}-\vec{q},\beta).
 \nonumber \\
\label{3.20} 
\end{eqnarray}
We observe that 
$c_{k\, L}(\vec{p},\alpha; \vec{p}-\vec{q},\beta)$ 
comprises the factor
$c_k(l_1 m_1,\, l_2 m_2)$ which is defined (compare with
expressions (\ref{3.9b}) and (\ref{3.9c})) by
\begin{eqnarray}
 c_{k}(l_1 m_1,\,l_2 m_2)=\int Y_{l_1}^{m_1\,*}(\Omega)\, 
 S_2^k(\Omega) \, Y_{l_2}^{m_2}(\Omega)\, d\Omega.
\label{3.21}
\end{eqnarray}
The quadrupolar contributions from the interaction
potentials (\ref{3.12a}), (\ref{3.12'a}) between 4$f$ electrons
and conduction electrons then read
\begin{mathletters}
\begin{eqnarray}
 \left. U^{fc}_{QQ} \right|_{inter}= 
 \sum_{\vec{q}}\, \rho_{k}^F(\vec{q})^{\dagger}\, 
 v_{k}^{F}{\,}_{k'}^{L}(\vec{q})\,
 \rho_{k'}^{L}(\vec{q}),
\label{3.22} 
\end{eqnarray}
and
\begin{eqnarray}
 \left. U^{fc}_{QQ} \right|_{intra}=
 C_{k}^{F}{\,}_{k}^{L}\,
 \sum_{\vec{q}}\, \rho_{k}^F(\vec{q})^{\dagger} \, 
 \rho_{k}^{L}(\vec{q}).
\label{3.23} 
\end{eqnarray}
\end{mathletters}
The contribution of the quadrupolar pair interaction
potential to expression (\ref{3.7}) is then given by
\begin{eqnarray}
 U_{QQ}^{fc}=\left. U^{fc}_{QQ} \right|_{inter} +
 \left. U^{fc}_{QQ} \right|_{intra}
\label{3.new23} 
\end{eqnarray}
The quadrupolar interactions between conduction electrons
follow from Eqs.~(\ref{3.18}) and (\ref{3.19a}):
\begin{mathletters}
\begin{eqnarray}
 \left. U^{cc}_{QQ} \right|_{inter}={\frac {1}{2}}
 \sum_{\vec{q}} \,  
 v_{k}^{L}{\,}_{k'}^{L'}(\vec{q})\;
 \eta \! \left(
 \rho_{k}^{L}(\vec{q})^{\dagger} 
 \rho_{k'}^{L'}(\vec{q})
 \right), 
\label{3.24}  
\end{eqnarray}
\begin{eqnarray}
 \left. U^{cc}_{QQ} \right|_{intra}= {\frac {1}{2}}
 C_{k}^{L}{\,}_{k}^{L'}\,
 \sum_{\vec{q}}\,  
 \eta \! \left(
 \rho_{k}^{L}(\vec{q})^{\dagger} 
 \rho_{k}^{L'}(\vec{q})
 \right).  
\label{3.25}  
\end{eqnarray}
\end{mathletters}
The contribution to expression (\ref{3.14}) is given by
\begin{eqnarray}
 U_{QQ}^{cc}=\left. U^{cc}_{QQ} \right|_{inter} +
 \left. U^{cc}_{QQ} \right|_{intra}
\label{3.new25} 
\end{eqnarray}
In order to select the contribution from conduction
electrons, we have studied the coefficients
$c_{k}(l_1 m_1,\,l_2 m_2)$. We find it convenient to use
as basis functions for the conduction electron states
the real spherical harmonics (see Sect.~\ref{sec:rd}).
Then the $s$ electron state is $|l=0,m=0\rangle $ while the
five $d$ electron states are $|2,m=0\rangle $; $|2,(m,s)\rangle $, 
$|2,(m,c)\rangle $, $m=1,2$. 
There are no transitions between 6$s$ states, the
transitions between 6$s$ and 5$d$ states are
\begin{eqnarray}
  \langle 0,0|S_2^1 |2,(1,s)\rangle  &=&
  \langle 0,0|S_2^2 |2,(1,c)\rangle  
    \nonumber \\
 &=& \langle 0,0|S_2^3 |2,(2,s)\rangle = \frac{1}{\sqrt{4\pi}}
 \label{3.26}
\end{eqnarray}
and zero otherwise.
The transition elements between 5$d$ states are quoted in Table~1.
%
\begin{table} 
\caption{
Calculated~coefficients~$c_k(l_1 m_1,\,l_2 m_2)$,
$k=1-3$, $l_1=l_2=2$; $m_i$ ($i=1,2$) stands
for the indices $(m_i,c)$ or $(m_i,s)$ of 
real spherical harmonics [25] 
Those functions which are not quoted here give zero
contributions.
\label{tab1}     } 
 
 \begin{tabular}{c c c c c c c} 
 ($m$) & ($m'$) & $\tau=(T_{2g},1)$ & $(T_{2g},2)$ &
 $(T_{2g},3)$ \\
\tableline
(0,c) & (1,c) & 0 & 0.09011 & 0 \\
(0,c) & (1,s) & 0.09011 & 0 & 0   \\
(0,c) & (2,s) & 0 & 0 & -0.18022   \\
(1,c) & (1,s) & 0 & 0 & 0.15608 \\
(1,c) & (2,c) & 0 & 0.15608 & 0  \\
(1,c) & (2,s) & 0.15608 & 0 & 0 \\
(1,s) & (2,c) & -0.15608 & 0 & 0 \\
(1,s) & (2,s) & 0 & 0.15608 & 0 
 \end{tabular} 
\end{table}

Since the coefficients $c_{k}(l_1 m_1,\,l_2 m_2)$ always occur in
conjunction with interaction matrix elements 
$v_{k}^{l_1 l_2}{\,}_{k'}^{{l'}_1 {l'}_2}$ \\ 
($v_{k}^{L}{\,}_{k'}^{L'}$),
we can immediately select the relevant matrix elements.
We observe that the indices $l_1$ and $l_2$
in the interaction matrix elements then directly refer to
$s$ or $d$ electrons and we will adopt the notation 
$v_k^{sd}{\,}_{k'}^{dd}$, $v_k^{ff}{\,}_{k'}^{sd}$ etc..
The indices $f$, $s$ and $d$ of these interaction matrix elements
refer only to the radial dependence, they are irrelevant
for symmetry considerations. 
The structure of the intersite quadrupole-quadrupole interaction
matrices are investigated in Appendix A.
From Eq.~(\ref{3.10})
it follows that the on-site
elements $C_k^{F}{\,}_{k'}^{L}$ are diagonal in
$k$, $k'$ and equal for $k$=1,2,3; the same holds for
$C_k^{L}{\,}_{k'}^{L'}$.
In the following we omit the indices $k$, $k'$ and write
$C^{ff\,sd}$, $C^{sd\,sd}$ etc.. Numerical values are given
in Table~2.
%
\begin{table} 
\caption{
Interaction parameters
$C^{A\,B}$,
$\lambda^{A\,B}$ and 
$\Lambda^{A\,B}$ 
calculated with $s-$, $d-$, $f-$ atomic radial distributions
and lattice constant $a$=9.753 a.u.($\gamma$-Ce);
$A=(l_1 l_2)$, $B=(l'_1 l'_2)$, $F=(ff)$.
\label{tab2}     } 
 
 \begin{tabular}{c c c c c} 
 $l_1 l_2$ & $l'_1 l'_2$ & $C$ (in K) & $\lambda$ (in K) 
               & $\Lambda$ (in K/{\AA}) \\
\tableline
 $ff$ & $ff$  &  -           & -2121  & -498   \\
 $sd$ & $ff$  & $\pm$4408   & -3924  & -950   \\
 $dd$ & $ff$  &  75389       & -6025  & -1459  \\
 $dd$ & $dd$  &  94559       & -17695 & -4283  \\
 $ds$ & $ds$  &  21297       & -7547  & -1822  \\
 $ds$ & $dd$  &  $\pm$34858  & -11545 & -2792 
 \end{tabular} 
\end{table} 
By introducing quadrupolar pair interactions on a same site
we generalize the concept of spherically symmetric on-site
electron-electron repulsions which is a characteristic property
of strongly correlated electron systems.

In order to treat the effect of conduction electrons on
the lattice contraction, we have studied the coupling
of quadrupole-quadrupole interactions with lattice
displacements within the tight-binding approach.
We start from expression $U_{QQT}$, Eq.~(\ref{2.15}),
and remind that $u_{\nu}(\vec{n})$ and $u_{\nu}(\vec{n}')$
refer to lattice displacements at different sites.
We first consider matrix elements of $V'_{\nu}$, 
expression (\ref{2.17}), between 4$f$ electron
and conduction electron states.
We proceed in analogy with Eqs.~(\ref{3.7})-(\ref{3.12b})
but retain only inter- site contributions.
We then find (compare with Eq. (5.13) of I)
in the long wavelength limit $\vec{q} \rightarrow 0$,
and $\vec{p}$ close to the star of $\vec{p}^X$:
\begin{eqnarray}
 U_{QQT}^{fc}=2i \sum_{\vec{p} \vec{q}} {\sum_{\nu(k)}}'
 v'_{\nu}{}_k^{F}{\,}_k^{L}(\vec{q},\vec{p})\,
 \rho_k^F(-\vec{p}) \, 
 \rho_k^{L}(\vec{p}) \, u_{\nu}(\vec{q}). 
\nonumber \\
 \label{3.28} 
\end{eqnarray}
Here the sum $\sum'$ refers to $\nu=x$, $y$ for $k$=3;
to $\nu=z$, $x$ for $k=2$ and to $\nu=z$, $y$ for $k=1$.
The coupling matrix is obtained as
\begin{eqnarray}
& & v'_{\nu}{}_3^{F}{\,}_3^{L}(\vec{q},\vec{p})=
 {\frac {1}{\sqrt{Nm}}} 
 \Lambda^{F\, L}  q_{\nu} \, a
 \cos\left({\frac{p_x a}{2}}\right) 
 \cos\left({\frac{p_y a}{2}}\right), 
 \nonumber \\
& & \label{3.29} 
\end{eqnarray}
with $\Lambda^{F\, L}$ given by 
$v'_{\nu}{}_k^{F}{}_{k}^{L}(\vec{n}-\vec{n}')$, 
where $\nu=x$ or $y$, $k=3$
and $\vec{X}(\vec{n})-\vec{X}(\vec{n}')=(a/2)(1,1,0)$
on the fcc lattice.
Here we have defined 
\begin{eqnarray}
 v'_{\nu}{}_k^{F}{\,}_{k'}^{L}(\vec{n}&-&\vec{n}')=
 \int\! dr\, r^2 \! \int\! dr'\, {r'}^2\, {\cal R}_f^2(r) \,
 \nonumber \\
 & \times &
 v'_{\nu\, k k'}(\vec{n}-\vec{n}'; r,r')\, 
 {\cal R}_{l_1}(r'){\cal R}_{l_2}(r'),
 \label{3.30} 
\end{eqnarray}
with $v'_{\nu\, k k'}(\vec{n}-\vec{n}'; r,r')$ given by Eq.~(\ref{2.16}).

The matrix elements of $V'_{\nu}$ between conduction electron
states are treated in analogy with Eqs.~(\ref{3.14})-(\ref{3.18}).
Now only inter- site terms occur.
In the limit $\vec{q} \rightarrow 0$ and $\vec{p}$ close
to the star of $\vec{p}^X$ we find
\begin{eqnarray}
 & & U_{QQT}^{cc}=i \sum_{\vec{p} \vec{q}} {\sum_{\nu(k)}}' 
 v'_{\nu}{}_k^{L}{\,}_k^{L'}(\vec{p},\vec{q}) \,
 \eta \! \left( \rho_k^{L}(-\vec{p}) \,
 \rho_k^{L'}(\vec{p}) \right) \, u_{\nu}(\vec{q}). 
 \nonumber \\
 & & \label{3.31} 
\end{eqnarray}
Here again we have the same relation between the indices
$\nu$ and $k$ as was the case for Eq.~(\ref{3.28}).
The coupling matrix reads
\begin{eqnarray}
  & & v'_{\nu}{}_3^{L} {\,}_3^{L'}(\vec{q},\vec{p})=
  {\frac {1}{\sqrt{Nm}}}
 \Lambda^{L\, L'} \, q_{\nu} \, a
 \cos\left({\frac{p_x a}{2}}\right) 
 \cos\left({\frac{p_y a}{2}}\right), 
 \nonumber \\
 & & \label{3.32} 
\end{eqnarray}
with $\Lambda^{L\, L'}$ given by 
$v'_{\nu}{}_k^{L}{\,}_{k}^{L'}(\vec{n}-\vec{n}')$, 
with $\nu=x$ or $y$, for $k=3$.
We finally quote the definition
\begin{eqnarray}
 v'_{\nu}{}_k^{L}{\,}_{k'}^{L'}(\vec{n}&-&\vec{n}')=
 \int\! dr\, r^2 \! \int \! dr'\, {r'}^2\, 
 {\cal R}_{l_1}(r){\cal R}_{l_2}(r) \,
 \nonumber \\
 &\times & v'_{\nu\, k k'}(\vec{n}-\vec{n}'; r,r')\, 
 {\cal R}_{l'_1}(r'){\cal R}_{l'_2}(r'),
 \label{3.33} 
\end{eqnarray}
where we use expression (\ref{2.16}).
Numerical values of $\Lambda^{L\, L'}$
are given in Table~2.

Taking into account the contributions due to conduction
electrons, we see that the interaction Hamiltonian (\ref{2.new29})
has to be replaced by
\begin{eqnarray}
 U=U_{TT}+U_{QQ}+U_{QQT} ,
 \label{3.new28} 
\end{eqnarray}
where
\begin{mathletters}
\begin{eqnarray}
 & &U_{QQ}=U_{QQ}^{ff}+U_{QQ}^{fc}+U_{QQ}^{cc},
 \label{3.new29a} \\ 
 & &U_{QQT}=U_{QQT}^{ff}+U_{QQT}^{fc}+U_{QQT}^{cc},
 \label{3.new29b}  
\end{eqnarray}
\end{mathletters}

Before studying the quadrupolar ordering and the accompanying 
lattice contraction (Sect.~\ref{sec:qo}) during the
$\gamma \rightarrow \alpha$ phase transition, we will next 
investigate the crystal field of $\gamma$-Ce in the presence of conduction
electrons.

\section {Crystal field of $\gamma$-Ce} 
\label{sec:cf}

In the disordered $\gamma$ phase there are only charge density
fluctuations of quadrupolar type and the quadrupolar Hamiltonian 
(\ref{3.new28}) (or (\ref{2.new29}))
averages to zero.
The first nontrivial orientational interaction then
corresponds to a crystal field Hamiltonian.
In Ref.~I we have defined the crystal field of $\gamma$-Ce
as the potential experienced by a single 4$f$ electron 
at a site $\vec{n}$
when spherically symmetric contributions ($l'$=0) from nuclei,
core electronic shells, conduction electrons, 4$f$
electrons at the twelve neighboring sites $\vec{n}'$ 
on the fcc lattice and 
similar terms from the homogeneous electronic density
in the interstitial regions are taken into account.
In I we have shown that crystal field effects are reduced
to a single particle term to which we have added
a spin-orbit coupling for the 4$f$ electron.
We first generalize the results of I by taking into account the
radial dependence of the 4$f$ electron density.
We start from Eq.~(\ref{2.8}) with
$\Lambda=(l=4,A_{1g}) \equiv \Lambda_1$ and 
$\Lambda'=(l=0,A_{1g}) \equiv 0$.  
Since we are dealing with a Coulomb potential
and a spherically symmetric charge distribution, 
the coupling function
$v_{\Lambda_1\, 0}(\vec{n},\vec{n}';r,r')$, Eq.~(\ref{2.3b}), does
not depend on $r'$ and we observe that Eq.~(\ref{2.6}) can
be written as
\begin{mathletters}
\begin{eqnarray}
 v_{\Lambda_1}^F{\,}_0^F(\vec{n}-\vec{n}')=
 v_{\Lambda_1}^F{}_0^{\bullet} \cdot Q_f,
\label{6.1a}
\end{eqnarray}
where
\begin{eqnarray}
  v_{\Lambda_1}^{F}{}_{0}^{\bullet} = \int dr\,r^2\, {\cal R}_f^2(r) \,
  v_{\Lambda_1\, 0}(\vec{n},\vec{n}'; r,r')
\label{6.1b}
\end{eqnarray}
is the same for all 12 neighbors. 
We obtain $v_{\Lambda_1}^{F}{}_{0}^{\bullet}<0$.
The charge in units $e$ of the 4$f$ electron 
at a neighboring site is given by
\begin{eqnarray}
  Q_f=\int dr'\,{r'}^2\,{\cal R}_f^2(r'). 
\label{6.1c}
\end{eqnarray}
\end{mathletters}
In our model $Q_f=1$. However, if we distinguish two regions
in the crystal, the first inside muffin-tin (MT) spheres and 
the second in the interstices, then in the MT-region $Q_f<1$. 
The other contributions to the crystal field are dealt with similarly. 
Since the interaction parameter $v_{\Lambda_1}^F{}_0^{\bullet}$ remains 
the same
it is only the charges $Q_i$ ($i$ stands for core, nucleus,
conduction electrons and interstitial contributions)
which we shall take care of.
Collecting the contributions from the various charges $Q_i$
together with $Q_f$, we obtain for the crystal field at site $\vec{n}$
\begin{mathletters}
\begin{eqnarray}
 V_{CF}^f(\vec{n})= B^f\,
  \rho_{\Lambda_1}^F(\vec{n}) ,  
  \label{6.2a}  
\end{eqnarray}
where $\Lambda_1 \equiv (l=4,A_{1g})$,
\begin{eqnarray}
  B^f = 
  {\frac {12}{\sqrt{4 \pi}}}\, Q_{eff}\,e\, v_{\Lambda_1}^{F}{}_0^{\bullet}  
  \label{6.2b}  
\end{eqnarray}
and
\begin{eqnarray}
 \rho_{\Lambda_1}^F(\vec{n})=\sum_{ij} c_{\Lambda_1}^F
  |i \rangle_{\vec{n}} \langle j|_{\vec{n}} .
\label{6.2c}
\end{eqnarray}
\end{mathletters}
Here $e$ refers to the electron charge at site $\vec{n}$ 
($e=-1$) and $Q_{eff}$ to the effective charge of the 
surrounding neighborhood.
From electrostatic considerations we find
(see Appendix~A of I) that
$Q_{eff}=(1+x_{int})Q_{MT}$, where $Q_{MT}$
is the total charge inside a MT-sphere (which is always positive)
and where the factor $x_{int} \approx 2.853$ accounts for
the charge contributions from interstices for touching
MT-spheres.
The coefficients $c_{\Lambda_1}^F$, Eq.~(\ref{6.2c}), are quoted in
Appendix A of I, they are diagonal in the basis (A.9-A.11) of I.
The expressions (\ref{6.2a})-(\ref{6.2c}) represent a refinement of
the crystal field calculations, Eqs.~(A.3)-(A.6) of I, where now
a radial distribution of the 4$f$ electron is taken into account.

In I we have calculated the crystal field coefficient $B^f$
(we used the notation $\Lambda$ for $B^f$) for
a fixed radius $r_f$. The obtained value $B^f$=346~K
corresponds to $r_f=1.156$~a.u. and not to
$r_f=1.378$~a.u. as was quoted erroneously in I.
With a radial distribution as specified in models
1), 2), and 3) of Sect.~\ref{sec:rd}, we obtain
$B^f=$970, 1403 and 1104~K, respectively.
This implies that the strength of the crystalline electric field 
of the 4$f$ electron
should be 2.8-4 times larger than calculated in Ref.~I
and therefore by the same factor (2.8-4) larger than 
the crystal field measured experimentally~\cite{Mil,Mur1}.
However, expression (\ref{6.2a}) is implicitly based on the
approximation that the distribution of conduction electrons
at site $\vec{n}$ is spherically symmetric and hence the
on-site interaction from $l=4$ multipoles is ignored.
In the following we will show that the effective crystal field
decreases if we take into account the average
on-site cubic distribution ($l=4$) of conduction electrons.

From Eqs.~(\ref{3.12'a},b) for the on-site interactions we observe 
that the 4$f$ electronic density $\rho_{\Lambda_1}^F$ couples
with the conduction electron density of the same symmetry 
($\Gamma=A_{1g}$, $l=4$).
In real space Eq.~(\ref{3.12'a}) 
with $\Lambda=\Lambda'=(l=4,A_{1g}) \equiv \Lambda_1$ reads
\begin{eqnarray}
   \left. U^{fc} \right|_{intra}=
    C^{F}_{\Lambda_1}{\,}^{L}_{\Lambda_1} 
    \sum_{\vec{n}} \rho_{\Lambda_1}^F(\vec{n})^{\dagger}\,
    \rho_{\Lambda_1}^{L}(\vec{n}) .
\label{6.3}
\end{eqnarray}
We obtain the crystal field potential due to 
$\rho_{\Lambda_1}^{L}(\vec{n})$, Eq. (\ref{3.11a}), by replacing this
quantity by its thermal average:
\begin{eqnarray}
  \langle \rho_{\Lambda_1}^{l_1 l_2} \rangle =
  {\frac {1}{N}} \sum_{\vec{k},\alpha}
  n_{\vec{k}}^{\alpha}\; 
  c_{\Lambda_1\,d\,d}(\vec{k},\alpha;\vec{k},\alpha)\;
  \delta_{l_1 2}\, \delta_{l_2 2}.
\label{6.4}
\end{eqnarray}
Here $n_{\vec{k}}^{\alpha}\; \delta_{\vec{q},0}\, \delta_{\alpha \beta}=
\langle a^{\dagger}_{\vec{k} \alpha} a_{\vec{k}-\vec{q}\, \beta} \rangle$ is
the Fermi distribution and $l_1=2$, $l_2=2$ takes into account
the fact that only $d$ electrons contribute to the matrix element
$c_{\Lambda_1\;l_1 l_2}$. We then write $\langle \rho^{c}_{\Lambda_1} \rangle=
\langle \rho^{d\, d}_{\Lambda_1} \rangle$.
The corresponding crystal field at site
$\vec{n}$ reads
\begin{mathletters}
\begin{eqnarray}
 V^{fc}_{CF}(\vec{n})=B^{fc}\, 
 \rho_{\Lambda_1}^F(\vec{n}) ,  
\label{6.5a} 
\end{eqnarray}
where
\begin{eqnarray}
 B^{fc}=C_{\Lambda_1}^{F} {\,}_{\Lambda_1}^{d d}\,
 \langle \rho^{c}_{\Lambda_1} \rangle ,  
\label{6.5b} 
\end{eqnarray}
\end{mathletters}
with $C_{\Lambda_1}^{F} {\,}_{\Lambda_1}^{d d}>0$. 
Collecting the contributions (\ref{6.2a}) and (\ref{6.5a})
we obtain the effective single particle potential acting on the 4$f$ electron
at site $\vec{n}$:
\begin{eqnarray}
   \tilde{V}_{CF}^f(\vec{n}) = \left[ B^f  +B^{fc} \right] \,
  \rho_{\Lambda_1}^F(\vec{n}) .  
\label{6.6} 
\end{eqnarray}
We recall that $B^f>0$.
The sign of $B^{fc}$ depends on $\langle \rho^{c}_{\Lambda_1} \rangle$.
We next investigate other interactions that affect 
$\langle \rho^{c}_{\Lambda_1} \rangle$.

We consider the crystal field acting on $\rho^{c}_{\Lambda_1}(\vec{n})$.
By studying the inter- site interactions we find that the matrix elements
$v_{\Lambda \Lambda'}(\vec{n},\vec{n}';r,r')$, Eq.~(\ref{2.3b}),
are negligibly small for $l=l'=4$.
On the other hand terms with $l=4$, $l'=0$ are significant and we
retain the interactions which involve
\begin{eqnarray}
  v_{\Lambda_1}^{d d}{\,}_{0}^{\bullet} = \int \! dr\,r^2\, {\cal R}_d^2(r) \,
  v_{\Lambda_1\, 0}(\vec{n},\vec{n'}; r,r') ,
\label{6.7}
\end{eqnarray}
where $v_{\Lambda_1}^{d d}{\,}_{0}^{\bullet}<0$ 
(compare with Eq.~(\ref{6.1b})).
Here again $v_{\Lambda_1}^{d d}{\,}_{0}^{\bullet}$ is independent of $r'$.
Proceeding in analogy with the calculation of $V^f_{CF}$,
we calculate the field due to the surrounding 4$f$ electrons,
core electrons, nuclei, conduction electrons and interstitial
contributions at site $\vec{n}$, thereby obtaining
\begin{mathletters}
\begin{eqnarray}
  V_{CF}^c(\vec{n}) = B^c \,
  \rho_{\Lambda_1}^c(\vec{n})  , 
\label{6.8a}  
\end{eqnarray}
where $\rho_{\Lambda_1}^c(\vec{n})=\rho_{\Lambda_1}^{dd}(\vec{n})$ and
\begin{eqnarray}
 B^c = 
  {\frac {12}{\sqrt{4 \pi}}} Q_{eff}\,e\, 
  v_{\Lambda_1}^{d d}{\,}_0^{\bullet},  
\label{6.8b}
\end{eqnarray}
\end{mathletters}
with $B^c=6093\;K>0$. The positive sign of the field $B^c$
implies that $\langle \rho^{c}_{\Lambda_1} \rangle$,
calculated with the crystal field $V_{CF}^c(\vec{n})$,
is negative.  Hence $B^{fc}$, Eq.~(\ref{6.5b}), is negative,
which leads to a reduction of the 
effective crystal field of a 4$f$ electron in Eq.~(\ref{6.6}).

The previous considerations indicate that the inclusion of
conduction electrons leads to a reduction of the crystal field
experienced by a 4$f$ electron.
However, a more rigorous approach should start with the
crystal field Hamiltonian of~$\gamma$-Ce,
\begin{eqnarray}
    U_0=U_0^c+U_{CF}+U_{so} .
\label{6.new9}
\end{eqnarray}
Here $U_0^c$ is the ``bare" electronic term (\ref{A.6}),
which includes the kinetic 
energy and the spherically symmetric part of the electronic
potential of conduction electrons,
$U_{so}=\sum_{\vec{n}} V_{so}({\vec{n}})$ stands for 
the spin-orbit couplings
of localized 4$f$ electrons (see Appendix A of I for details).
$U_{CF}$ is the crystal field 
comprising 4$f$ electrons and conduction electrons: 
\begin{mathletters}
\begin{eqnarray}
   & &U_{CF} = \sum_{\vec{n}} V_{CF}(\vec{n}) , \label{6.9a} \\
   & &V_{CF}(\vec{n}) = V_{CF}^f(\vec{n}) + V_{CF}^{fc}(\vec{n}) 
  + V_{CF}^c(\vec{n}) +V_{CF}^{cc}(\vec{n}) . \nonumber \\
\label{6.9b}
\end{eqnarray}
\end{mathletters}
Here
\begin{eqnarray}
   V_{CF}^{cc}(\vec{n}) =
    C^{d\,d}_{\Lambda_1}{\,}^{d\,d}_{\Lambda_1} \;
    \rho_{\Lambda_1}^{d\,d}(\vec{n})^{\dagger}\,
    \rho_{\Lambda_1}^{d\,d}(\vec{n}) 
\label{6.10}
\end{eqnarray}
is the on-site $l=4$, $l'=4$ interaction between
conduction electrons.
In a mean-field approximation $U_{CF}$
leads to self-consistent crystal field potentials 
$\tilde{V}_{CF}^c(\vec{n})$ and $\tilde{V}_{CF}^f(\vec{n})$ 
for conduction and 4$f$ electrons, respectively.
A quantitative calculation of these effects 
is beyond the scope of the present work.

\section {Quadrupolar ordering} 
\label{sec:qo}

In Section~\ref{sec:tb} we have found that the
system of localized 4$f$ electrons and of conduction
electrons are coupled by means of the intra- site and
inter- site quadrupolar potentials (\ref{3.23}) and (\ref{3.22}).
Such a bilinear coupling suggests that an ordering of
4$f$ electron density quadrupoles should imply an
ordering of conduction electron quadrupoles and
vice versa. 
In Appendix A we have investigated the wave vector dependence of
the quadrupole-quadrupole interaction matrices and the condition for
quadrupolar order.
The present section is divided into two parts. In part A we study
the interplay of quadrupolar ordering of 4$f$ electrons and
conduction electrons on a rigid cubic lattice; in part B,
where we consider a deformable lattice, we show that quadrupolar
ordering implies a lattice contraction.

\noindent
{\Large {\bf A}}. 
We are investigating the possibility of a condensation of
quadrupolar densities in a $Pa{\bar 3}$ structure.
At the $X$-point of the BZ, the 
matrices of inter- site interactions ($v_k^{F}{\,}_{k'}^{L}(\vec{q})$)
and ($v_k^{L}{\,}_{k'}^{L'}(\vec{q})$) become diagonal in 
$k$, $k'$ and have two degenerate negative eigenvalues.
We then are led in analogy with the condensation scheme (\ref{2.14a},b)
for 4$f$ electrons to suggest the condensation scheme
$Fm{\bar 3}m \rightarrow Pa{\bar 3}$ for the quadrupole densities of
conduction electrons:
\begin{mathletters}
\begin{eqnarray}
 & & {\bar \rho}_3^{L}(\vec{q}_x^X)=
 {\bar \rho}_1^{L}(\vec{q}_y^X)=
 {\bar \rho}_2^{L}(\vec{q}_z^X)
 ={\bar \rho}^{L} \sqrt{N} \neq 0;
 \label{3.27a} \\ 
 & & {\bar \rho}_2^{L}(\vec{q}_x^X)=
 {\bar \rho}_3^{L}(\vec{q}_y^X)=
 {\bar \rho}_1^{L}(\vec{q}_z^X)=0,
 \label{3.27b} 
\end{eqnarray}
\end{mathletters}
where $L=(sd)$, ($ds$), ($dd$).

We disentangle the various contributions to $U_{QQ}$,  
Eq. (\ref{3.new29a}), in the ordered $\alpha$ phase 
where we assume a simultaneous condensation
of quadrupolar densities of 4$f$ electrons and conduction electrons.
Taking into account the condensation scheme (\ref{2.14a},b),
we obtain from Eq.~(\ref{2.12a})
\begin{eqnarray}
 {\frac {1}{N}} U_{QQ}^{ff} = {\frac {3}{2}} \lambda^{F F}
 \left( {\bar \rho}^F \right)^2,
 \label{4.1} 
\end{eqnarray}
where $\lambda^{F F}$ stands for $\lambda_{X_5^+}$,
the twofold degenerate negative eigenvalue of
$v^{F F}(\vec{q}^X)$ (see Appendix A). 
Similarly, using in addition the condensation
scheme (\ref{3.27a},b), we obtain from Eq.~(\ref{3.23})
\begin{mathletters}
\begin{eqnarray}
 {\frac {1}{N}} \left. U_{QQ}^{fc} \right|_{intra}=
 3\left[ 2C^{F\,ds} {\bar \rho}^F {\bar \rho}^{ds}+
 C^{F\,dd} {\bar \rho}^F {\bar \rho}^{dd} \right],
 \label{4.2a} 
\end{eqnarray}
and from Eq.~(\ref{3.22})
\begin{eqnarray}
 {\frac {1}{N}} \left. U_{QQ}^{fc} \right|_{inter}=
  3 \left[ 2\lambda^{F\,ds} {\bar \rho}^F {\bar \rho}^{ds}+
 \lambda^{F\,dd} {\bar \rho}^F {\bar \rho}^{dd} \right],
 \label{4.2b} 
\end{eqnarray}
\end{mathletters}
where $\lambda^{F\,ds}$ and $\lambda^{F\,dd}$ are 
the twofold degenerate negative eigenvalues of the matrices
$v^{F\, ds}(\vec{q}^X)$ and $v^{F\, dd}(\vec{q}^X)$, respectively
(see Appendix A).
Finally expressions (\ref{3.25}) and (\ref{3.24}) lead to
\begin{mathletters}
\begin{eqnarray}
 {\frac {1}{N}} \left. U_{QQ}^{cc} \right|_{intra}=
 {\frac {3}{2}}
 \left[ 4C^{sd\,sd} ({\bar \rho}^{sd})^2 +
 4C^{sd\,dd} {\bar \rho}^{sd} {\bar \rho}^{dd} \right.
 \nonumber \\ 
 \left. +
 C^{dd\,dd} ({\bar \rho}^{dd})^2  \right],
 \label{4.3a} 
\end{eqnarray}
and 
\begin{eqnarray}
 {\frac {1}{N}} \left. U_{QQ}^{cc} \right|_{inter}=
 {\frac{3}{2}}\left[ 4\lambda^{sd\,sd} ({\bar \rho}^{sd})^2 +
 4\lambda^{sd\,dd} {\bar \rho}^{sd} {\bar \rho}^{dd} \right.
 \nonumber \\
\left.  +\lambda^{dd\,dd} ({\bar \rho}^{dd})^2  \right].
 \label{4.3b}
\end{eqnarray}
\end{mathletters}
We observe that all coefficients $C^{L\, L'}$
are positive while $\lambda^{L\, L'}$ are
negative (see Table~2).
Obviously the inter-site interaction (\ref{4.3b}) favors
quadrupolar order while the intra- site coupling (\ref{4.3a})
disfavors quadrupolar order of conduction electrons.

The leading quadrupolar interaction in the ordered $\alpha$ phase,
\begin{eqnarray}
 U_{QQ} = U_{QQ}^{ff} +
  \left. U_{QQ}^{fc} \right|_{intra} +
  \left. U_{QQ}^{fc} \right|_{inter}  \nonumber \\
  +\left. U_{QQ}^{cc} \right|_{intra} +
  \left. U_{QQ}^{cc} \right|_{inter} ,
\label{4.4}
\end{eqnarray}
is a quadratic form in ${\bar \rho}^F$, ${\bar \rho}^{sd}$
and ${\bar \rho}^{dd}$, in particular the terms (\ref{4.2a})
and (\ref{4.2b}) represent a bilinear coupling between
the localized 4$f$ quadrupolar density ${\bar \rho}^F$ and
the conduction electron quadrupolar densities ${\bar \rho}^{dd}$
and ${\bar \rho}^{sd}$.
Since in tight-binding the $(6s5d)^3$ electrons are hybridized,
we introduce the total conduction electron density
\begin{eqnarray}
 {\bar \rho}^c={\bar \rho}^{dd}+2{\bar \rho}^{sd}.
\label{4.5}
\end{eqnarray}
Defining average interaction coefficients
\begin{mathletters}
\begin{eqnarray}
 & &C^{fc}={\frac {1}{2}}(C^{F\,ds}+C^{F\,dd}), \label{4.6a} \\
 & &\lambda^{fc}={\frac {1}{2}}(\lambda^{F\,ds}+
         \lambda^{F\,dd}), \label{4.6b} \\
 & &C^{cc}={\frac {1}{3}}(C^{ds\,ds}+C^{dd\,dd}+C^{ds\,dd}), \label{4.6c} \\
 & &\lambda^{cc}={\frac {1}{3}}(\lambda^{ds\,ds}+\lambda^{dd\,dd}+
 \lambda^{ds\,dd}),     \label{4.6d}
\end{eqnarray}
\end{mathletters}
we approximate $U_{QQ}$ by an effective interaction
\begin{eqnarray}
 {\frac{1}{N}}U_{QQ} \approx
 {\frac {3}{2}} \left[ 
 \lambda^{F F}({\bar \rho}^F)^2
 +2 A^{fc} {\bar \rho}^F {\bar \rho}^c
 +  A^{cc} ({\bar \rho}^c)^2 \right] ,
\label{4.7}
\end{eqnarray}
where
\begin{mathletters}
\begin{eqnarray}
 A^{fc}=C^{fc} + \lambda^{fc}
\label{4.8a}
\end{eqnarray}
and
\begin{eqnarray}
 A^{cc}=C^{cc} + \lambda^{cc}.
\label{4.8b}
\end{eqnarray}
\end{mathletters}
We observe that $A^{fc}>0$, $A^{cc}>0$.
For a fixed value of ${\bar \rho}^F$, we minimize $U_{QQ}$
with respect to ${\bar \rho}^c$ and obtain
\begin{eqnarray}
  {\bar \rho}^c= - {\frac {A^{fc}}{A^{cc}}}{\bar \rho}^F.
\label{4.9}
\end{eqnarray}
We see that quadrupolar order ${\bar \rho}^F$ of 4$f$
electrons produces (as a type of mirror image)
a quadrupolar order of conduction electrons ${\bar \rho}^c$
of opposite sign.
Herewith we associate the pictorial representation
of Fig.~3 and 4. 
%
\begin{figure} 
\centerline{
\epsfig{file=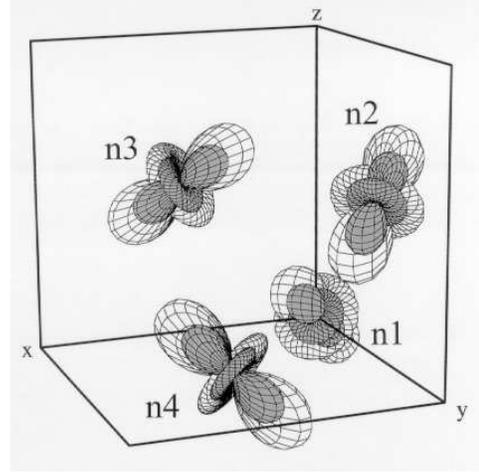,width=0.35\textwidth}
} 
\vspace{0.5cm}
\caption{
$Pa{\bar 3}$ structure of the ordered $\alpha$ phase
with 4 sublattices ($n1-n4$).
Grey quadrupoles correspond to inner 4$f$ electron
densities, white quadrupoles - to outer conduction
electron densities with the opposite sign as shown
in figure 4.
} 
\label{fig3} 
\end{figure} 
%
\begin{figure} 
\centerline{
\epsfig{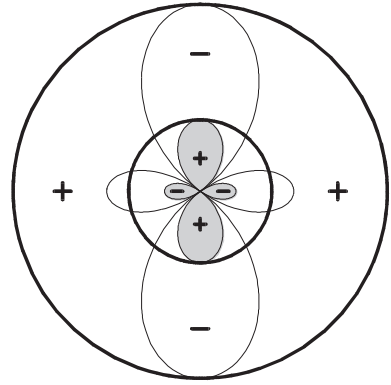}
} 
\vspace{0.5cm}
\caption{
Quadrupolar density distribution of the 4$f$ electron
(inside sphere of $r$=1.3 a.u.) and of one conduction electron
(spherical radius $r$=3.4 a.u.), on scale.
} 
\label{fig4} 
\end{figure} 
Regions of an excess density ($+$) of the 4$f$ electron
distribution overlap with a depletion ($-$) in density of
the conduction electrons and vice versa (Fig.~4). 
Substituting the right hand side of Eq.~(\ref{4.9})
into (\ref{4.7}) we get
\begin{eqnarray}
  {\frac {1}{N}} U_{QQ} = - {\frac {3}{2}}
  \left[ |\lambda^{F F}|+{\frac{(A^{fc})^2}{A^{cc}}} \right]
  ({\bar \rho}^F)^2
\label{4.10}
\end{eqnarray}
where we use the fact that $\lambda^{F F}<0$.
The bilinear coupling between quadrupolar 
densities of 4$f$ electrons and conduction electrons leads
to an increase of the attractive interaction between 4$f$ quadrupolar
densities.

The results of the last two sections can
be summarized by suggesting an effective Hamiltonian in the
disordered $\gamma$ phase,
\begin{eqnarray}
   H = U_{QQ}+U_0 , \label{4.new12a} 
\end{eqnarray}
where $U_0$ is the crystal field Hamiltonian, (\ref{6.new9}), and 
\begin{eqnarray}
   U_{QQ} = U_{QQ}^{ff}+U_{QQ}^{fc}+U_{QQ}^{cc} .
\end{eqnarray}
Here
\begin{mathletters}
\begin{eqnarray}
 U_{QQ}^{ff} &=& -\frac{1}{2}|\lambda^{ff}| \, 
 ( \rho_2^f(\vec{q}^X_x)^{\dagger} \rho_2^f(\vec{q}^X_x) + 
 \rho_3^f(\vec{q}^X_x)^{\dagger} \rho_3^f(\vec{q}^X_x) )
 \nonumber \\ 
  & &+  c.p. , \label{4.new12'a} 
\end{eqnarray}
\begin{eqnarray}
 & &U_{QQ}^{fc}= A^{fc} \sum_{\vec{q}} \sum_k
 \rho_k^f(\vec{q})^{\dagger} \rho_k^c(\vec{q}),
\label{4.new12'b} \\
 & &U_{QQ}^{cc}= \frac{1}{2} A^{cc} \sum_{\vec{q}} \sum_k
 \rho_k^c(\vec{q})^{\dagger} \rho_k^c(\vec{q}) ,
\label{4.new12'c} 
\end{eqnarray}
\end{mathletters}
where $\rho_k^f(\vec{q})$, $k=1-3$, is the quadrupolar density
operator (\ref{2.10a}) of 4$f$ electrons while $\rho_k^c(\vec{q})$
is the quadrupolar density of conduction electrons (\ref{3.20})
with $\rho_k^c=\rho_k^{dd}+2\rho_k^{sd}$.
The first term, Eq.~(\ref{4.new12'a}), represents an effective
attraction at the $X$ point of the Brillouin zone.
The other terms in (\ref{4.new12'a}) ($c.p.$) follow by cyclic
permutation of indices, {\it i.e.} when $\vec{q}^X_x \rightarrow \vec{q}^X_y$,
$k$ changes $2 \rightarrow 3$ and $3 \rightarrow 1$, etc.
The direct lattice is face centered cubic. In that case
the condensation schemes (\ref{2.14a},b) and (\ref{3.27a},b)
correspond to a structural phase transition where the space
group changes from $Fm{\bar 3}m$ to $Pa{\bar 3}$.

Starting from the interaction (\ref{4.7}) we construct
a Landau free energy in the condensed phase.
We take ${\bar \rho}_1 \equiv {\bar \rho}^F$ and 
${\bar \rho}_2 \equiv {\bar \rho}^c$ as components
of a two dimensional vector ${\bar \rho}=({\bar \rho}_1,{\bar \rho}_2)$
and define the matrix
\begin{eqnarray}
 J =
 \left[ \begin{array}{c c}
 -|\lambda^{F F}| & A^{fc} \\
 A^{fc} & A^{cc}
 \end{array} \right] .  
\label{4.new14} 
\end{eqnarray}
Inspired from the theory of orientational order in molecular
crystals we write
\begin{eqnarray}
 {\frac {{\cal F}}{N}} = {\frac {{\cal F}_0}{N}} + {\frac {3}{2}}
 \sum_{ij} {\bar \rho}_i\, \chi^{-1}_{ij}\, {\bar \rho}_j
  +{\cal F}^{(3)}+{\cal F}^{(4)}.
\label{4.new15}
\end{eqnarray}
Here ${\cal F}_0$ is the free energy in the disordered phase:
\begin{eqnarray}
  {\cal F}_0=-T \ln Tr [e^{-U_0/T}] ,
\label{4.new15a}
\end{eqnarray}
where $U_0$ is the crystal field, Eq.~(\ref{6.new9}).
The quantity
\begin{eqnarray}
  \chi^{-1} = \left[ T \langle \rho\, \rho \rangle_0^{-1} + J \right]
\label{4.new16}
\end{eqnarray}
is the inverse susceptibility matrix, $T$ is the temperature and
\begin{eqnarray}
 \langle \rho\, \rho \rangle_0 =
 \left[ \begin{array}{c c}
 \langle (\rho_k^F(\vec{n}))^2 \rangle_0 & 
 0  \\
 0 & 
 \langle (\rho_k^c(\vec{n}))^2 \rangle_0
 \end{array} \right] .  
\label{4.new17} 
\end{eqnarray}
Here $\langle ... \rangle_0$ are single particle thermal
expectation values that have to be calculated by means
of $U_0$, Eq.~(\ref{6.new9}). Cubic symmetry implies 
(see Appendix B for details) that the
three quadrupolar components are equal and for
expectation values of conduction electrons we obtain:
\begin{eqnarray}
\langle \rho_k^{L}(\vec{n}) \rho_{k'}^{L'}(\vec{n}) \rangle_0
  = {\frac {1}{N^2}} \sum_{\alpha\, \beta} \sum_{\vec{p}\, \vec{h}}
 (1-n_{\vec{p}\, \beta})\, n_{\vec{h}\, \alpha}\; 
 \nonumber \\ \times \,
 \left| c_{k\, L}(\vec{h}, \alpha; \vec{p},\beta)
 \right|^2 
 \delta_{L\, L'}\, \delta_{k\, k'} .
\label{B.7} 
\end{eqnarray}
Here $\delta_{L\, L'}=\delta_{l_1\,l'_1} \delta_{l_2\,l'_2}$ .
The contributions ${\cal F}^{(3)}$ and ${\cal F}^{(4)}$
stand for the third and fourth order terms in ${\bar \rho}$.
Symmetry of the order parameter components
(see condensation schemes (\ref{2.14a},b) and (\ref{3.27a},b)) implies
that there exists a non-zero third order cubic invariant ${\cal F}^{(3)}$
and hence the transition $Fm{\bar 3}m \rightarrow Pa{\bar 3}$ 
is of first order.
As a first approximation we investigate the possibility of a second order
phase transition. Neglecting ${\cal F}^{(3)}$ and ${\cal F}^{(4)}$, 
we minimize ${\cal F}$ with respect to ${\bar \rho}^F$ and to ${\bar \rho}^c$.
We obtain two coupled homogeneous equations
from which it follows again that ${\bar \rho}^F$ and ${\bar \rho}^c$
have to be of opposite sign.
The compatibility condition leads to the transition temperature
\begin{eqnarray}
 T_C= {\frac {1}{2}} \left(
 J_{ff}-J_{cc} + \sqrt{(J_{ff}+J_{cc})^2+4(J_{fc})^2} \right),
\label{4.new18}
\end{eqnarray}
where
\begin{mathletters}
\begin{eqnarray}
 & &J_{ff}= |\lambda^{FF}|\, \langle (\rho_k^F(\vec{n}))^2 \rangle_0 , 
 \;\;\;
 J_{cc}=A^{cc} \, \langle (\rho_k^c(\vec{n}))^2 \rangle_0 , \label{4.new19a} \\
 & &J_{fc}=A^{fc} \, \sqrt{\langle (\rho_k^F(\vec{n}))^2 \rangle_0
 \langle (\rho_k^c(\vec{n}))^2 \rangle_0 } .
\label{4.new19b}
\end{eqnarray}
\end{mathletters}
We observe that $T_C>J_{ff}$ and the inclusion of conduction
electrons leads to an increase of transition temperature.
The calculation of the first order transition temperature $T_1$,
($T_1>T_C$), and of the accompanying discontinuities
${\bar \rho}^c(T_1)$ and ${\bar \rho}^F(T_1)$ in the
coupled order parameters would require the evaluation of the
higher order terms ${\cal F}^{(3)}$ and ${\cal F}^{(4)}$.
Such an endeavor poses not only analytical but also
numerical problems that are beyond the scope of the present work.
We therefore will adopt below (Sect.~6) an alternative point of view and
suggest experiments in order to check qualitatively the
predictions of the theory.

\noindent
{\Large {\bf B}}. 
We next investigate the lattice strains at the transition to the
$Pa{\bar 3}$ phase. 
We recall the result $U_{QQT}^{ff}$, Eq.~(\ref{2.new27}),
for the coupling of strains to 4$f$ electrons with
quadrupolar order.
The same procedure, as outlined in Sect.~2, is now applied to
derive the coupling of longitudinal strains to conduction
electrons.
Using the condensation schemes (\ref{2.14a},b) and (\ref{3.27a},b)
and taking into account the definitions of
interaction parameters $\Lambda$ (Table~2),
we obtain from Eqs.~(\ref{3.28}) and (\ref{3.29})
\begin{eqnarray}
 & & {\frac {1}{N}}U_{QQT}^{fc}=-4a \left[
 2 \Lambda^{F\,ds}{\bar \rho}^F {\bar \rho}^{ds} + 
 \Lambda^{F\,dd}{\bar \rho}^F {\bar \rho}^{dd} \right]
 \sum_{\nu} \epsilon_{\nu \nu} .
 \nonumber \\
 & & \label{4.21}
\end{eqnarray}
Similarly we get from from Eqs.~(\ref{3.31}) and (\ref{3.32})
\begin{eqnarray}
 {\frac {1}{N}}U_{QQT}^{cc}=-2a \left[
 4 \Lambda^{ds\,ds} \left( {\bar \rho}^{ds} \right)^2 + 
 4 \Lambda^{ds\,dd}{\bar \rho}^{ds} {\bar \rho}^{dd} 
 \right. \nonumber \\  \left. 
   +\Lambda^{dd\,dd} \left( {\bar \rho}^{dd} \right)^2
 \right]
 \sum_{\nu} \epsilon_{\nu \nu}.
\label{4.22}
\end{eqnarray}
The coefficients $\Lambda$ are quoted in Table 2.
We define the total interaction potential of ordered
quadrupolar electron densities coupled to longitudinal strains by
\begin{eqnarray}
 U_{QQT} = U_{QQT}^{ff}+U_{QQT}^{fc}+U_{QQT}^{cc}.
\label{4.23}
\end{eqnarray}
Here the right hand side terms are given by Eqs.~(\ref{2.new27}),
(\ref{4.21}) and (\ref{4.22}), respectively.
In analogy with expressions (\ref{4.6b},d) we define
\begin{mathletters}
\begin{eqnarray}
 & & \Lambda^{fc}= {\frac {1}{2}}
 (\Lambda^{F\,ds}+\Lambda^{F\,dd}),
    \label{4.24a} \\
 & & \Lambda^{cc}= {\frac {1}{3}}
 (\Lambda^{ds\,ds}+\Lambda^{dd\,dd}+\Lambda^{ds\,dd}),
    \label{4.24b} 
\end{eqnarray}
\end{mathletters}
and use again definition (\ref{4.5}).
Then we approximate 
\\ 
$U_{QQT}/N$ by an effective interaction
\begin{eqnarray}
 {\frac {1}{N}}U_{QQT} \approx -2a
  [ \Lambda^{FF} ( {\bar \rho}^F )^2 + 
 2 \Lambda^{fc} {\bar \rho}^F {\bar \rho}^c &+& 
 \Lambda^{cc}  ( {\bar \rho}^c )^2 ]
 \nonumber \\
 & \times &  \sum_{\nu} \epsilon_{\nu \nu}.
\label{4.25}
\end{eqnarray}
This relation allows us to express the longitudinal 
strains as function of the order parameters.
The contributions $U_{TT}$ and $U_{QQT}$ lead to
a supplementary term in the free energy 
(compare with Eq.~(\ref{4.new15})):
\begin{eqnarray}
 \frac{1}{N} {\cal F}_{QQT}=\frac{1}{N} U_{TT}
 +\frac{1}{N} U_{QQT} .
\label{5.new26}
\end{eqnarray}
(There are still additional contributions from
thermal lattice vibrations, but these are irrelevant here.)
For a given quadrupolar order, 
{\it i.e.} ${\bar \rho}^F$ and ${\bar \rho}^c$ fixed, we minimize \\ 
${\cal F}_{QQT}[{\bar \rho}^F,{\bar \rho}^c,\epsilon_{\nu \nu}]$ with respect 
to the strains $\epsilon_{xx}$,
$\epsilon_{yy}$ and $\epsilon_{zz}$ and obtain
\begin{eqnarray}
 \epsilon_{xx} = \epsilon_{yy} &=& \epsilon_{zz}=
 8a^{-2} \kappa_L \,
 \nonumber \\
 & \times & \left[ \Lambda^{F F} ( {\bar \rho}^F )^2 + 
 2 \Lambda^{fc} {\bar \rho}^F {\bar \rho}^c + 
 \Lambda^{cc}  ( {\bar \rho}^c )^2  \right],
\label{4.26}
\end{eqnarray}
where $a$ is the cubic lattice constant and where $\kappa_L$
is the bare linear compressibility.
From the numerical values of Table~2 we obtain
$\Lambda^{F F}=-498$, $\Lambda^{fc}=-1205$, $\Lambda^{cc}=-2966$
(units K/{\AA}). Since ${\bar \rho}^F$ and ${\bar \rho}^c$
are of opposite sign, the $fc$ contribution on the right hand
side of expression (\ref{4.26}) leads to an expansion of the
lattice while the $ff$ and $cc$ contributions lead to a
contraction.
From Table~2 we also find that $A^{fc}=30516$~K, $A^{cc}=24033$~K,
and hence from Eq.~(\ref{4.9}) we deduce that 
$|{\bar \rho}^c|>|{\bar \rho}^F|$.
Hence we conclude that inclusion of the conduction electrons in
expression (\ref{4.26}) leads to an increase of the lattice contraction
by a factor 4.4 in comparison with the effect of the 4$f$ electrons.
A numerical calculation of the lattice contraction
$\triangle a= a \epsilon_{xx}$ at the first order phase transition
would require the knowledge of the discontinuities 
${\bar \rho}^c(T_1)$ and ${\bar \rho}^F(T_1)$.
Experimentally \cite{Kos} one finds in the pressure $(P)$ - temperature  phase
diagram of the $\gamma-\alpha$ transition in Ce
indications of a critical end point where the lattice contraction
vanishes.
Although the transition $Fm{\bar 3}m \rightarrow Pa{\bar 3}$
is always of first order and hence leads to finite
discontinuities of ${\bar \rho}^c(T_1)$ and ${\bar \rho}^F(T_1)$,
there exists the possibility that the relative importance
of the coefficients $\Lambda^{fc}$ and $\Lambda^{cc}$,
which have opposite sign, changes as a function of $(P,T)$
and hence the expression within square brackets on the right hand
side of Eq.~(\ref{4.26}) could vanish or even change sign without
the requirement that ${\bar \rho}^c(T_1)=0$ and ${\bar \rho}^F(T_1)=0$.
This opens up a possibility of a lattice expansion during the discussed
first order phase transition. 
Recent experiments on YbInCu$_4$ which exhibits an isostructural
phase transition similar to the $\gamma-\alpha$ change observed in cerium 
(see \cite{Sar} for references) indicate a 0.5\% volume
expansion at the transition. Within the present theory such
behavior can be understood if for Yb in YbInCu$_4$
the $fc$ term responsible for expansion prevails over $ff$ and $cc$ 
contributions, Eq.~(\ref{4.26}).

%
%
\section {Discussion} 
\label{sec:con} 

The present paper is an extension of our previous 
model of the $\gamma-\alpha$ phase transition in Ce (Ref.~\cite{NM} or I).
We predict that the $\gamma-\alpha$ transition is 
accompanied by a symmetry change $Fm{\bar 3}m \rightarrow Pa{\bar 3}$
in the electronic structure.
The idea of such a proposal is borrowed from the theory of
molecular crystals, where $Pa{\bar 3}$ structures due to orientational
order of molecular mass distribution are not unusual.
For example, such crystal symmetry occurs in NaO$_2$ \cite{Zel}, 
N$_2$ \cite{Sco} and in solid C$_{60}$ \cite{Dav}. 
The conventional characterization of the $\gamma-\alpha$ transition in Ce
as a phase transition ``without change of symmetry" 
is based on several X-ray
diffraction experiments \cite{Kos}.
It is possible that domain formation in the $\alpha$-phase
has precluded an identification of this phase as
a $Pa{\bar 3}$ structure.
If new experiments are done, particular attention
should be given to the possible coexistence
of domains \cite{Par}.

In comparison with our previous work, Ref.~I,
we have extended the model in two respects. First, we have included
quadrupolar interactions between 4$f$ and conduction electrons
and secondly, we have calculated the relevant parameters of
interactions (Table 2 and Sect.~2) by using the radial dependences of 
valence electrons
obtained from a DFT-LDA calculation of a cerium atom.
To our knowledge, in the literature there exists no microscopic derivation
of multipolar interactions between conduction and localized
electrons in solids although the concept of quadrupolar moment
of a $4f$ electron shell is well established \cite{Mor1,Mor2}. 
Therefore in Sect.~3 we have presented
a detailed calculation of multipolar interactions 
treating band conduction electrons in second quantization with
wave functions in tight-binding approximation.
Since the intersite quadrupole-quadrupole interaction is short ranged
and anisotropic,
special attention has been given to lattice site symmetry.
While we have restricted ourselves here
(see in particular Appendix A) to the case of an fcc lattice
($\gamma \rightarrow \alpha$ Ce), our procedure is general
and can easily be extended to other structures.

Our results can be understood on the basis
of the following generalized
Hamiltonian which we ascribe to the $\gamma$ phase:
\begin{eqnarray}
 H_{\gamma}= U_0+U_{QQ}+U_{QQT}+U_{TT} .
\label{7.1}
\end{eqnarray}
Here $U_0$ is the crystal field Hamiltonian (\ref{6.new9}).
The term $U_{QQ}$ represents the quadrupole-quadrupole
interaction comprising contributions from $4f$ and conduction
electrons (Eq. (\ref{3.new29a})).
The quadrupolar interaction between localized electrons
is due only to intersite contributions on the fcc lattice,
see Eq. (\ref{2.12a}) in Fourier space. The presence
of conduction electrons leads to two types of contributions: 
inter-site terms, given by (\ref{3.22}) and (\ref{3.24}) for
$U_{QQ}^{fc}$ and $U_{QQ}^{cc}$, respectively, and on-site
terms, given by (\ref{3.23}) and (\ref{3.25}), correspondingly.
Collecting inter- site and on- site terms separately,
we write
\begin{eqnarray}
 U_{QQ} =\left. U_{QQ} \right|_{inter} +
 \left. U_{QQ} \right|_{intra},
\label{7.2} 
\end{eqnarray}
where
\begin{mathletters}
\begin{eqnarray}
 & &\left. U_{QQ} \right|_{inter}=\sum_{\vec{q}} \sum_{k,k'} \,  
 \left( {\frac {1}{2}}
  v_{k}^{F}{\,}_{k'}^{F}(\vec{q})\;
 \rho_{k}^{F}(\vec{q})^{\dagger} 
 \rho_{k'}^{F}(\vec{q}) \right.  \nonumber \\
 & & \left. + v_{k}^{F}{\,}_{k'}^{L}(\vec{q})\, 
 \rho_{k}^{F}(\vec{q})^{\dagger} 
 \rho_{k'}^{L}(\vec{q}) + \!
 {\frac {1}{2}}
  v_{k}^{L}{\,}_{k'}^{L'}(\vec{q})\;
 \eta \! \left(
 \rho_{k}^{L}(\vec{q})^{\dagger} 
 \rho_{k'}^{L'}(\vec{q})
 \right) \right) ,  
 \nonumber \\
\label{7.3a}  
\end{eqnarray}
and the summation over $L(L')=(sd)$, $(ds)$, $(dd)$ is implied.
The structure of the interaction matrices $v^{A B}(\vec{q})$,
$A(B)=F$ or $L$, is discussed in Appendix A, the relevant
eigenvalues $\lambda^{A B}$ at $\vec{q}^X$ are quoted in Table 2.
The intra- site contributions are given by
\begin{eqnarray}
\left. U_{QQ} \right|_{intra}= 
 \sum_{\vec{q}} \sum_k  & &
 \left( 
 C^{F L}\,
 \rho_{k}^{F}(\vec{q})^{\dagger} 
 \rho_{k}^{L}(\vec{q}) \right.  \nonumber \\  
 & & \left. + {\frac {1}{2}} C^{L L'}\,
 \eta \! \left(
 \rho_{k}^{L}(\vec{q})^{\dagger} 
 \rho_{k}^{L'}(\vec{q})
 \right) \right) ,  
\label{7.3b}  
\end{eqnarray}
\end{mathletters}
where the calculated parameters $C^{A B}$ are quoted in Table 2.
Next in Eq. (\ref{7.1}), $U_{QQT}$, 
given by (\ref{3.new29b}), is a correction
to $U_{QQ}$ for a deformable lattice, while $U_{TT}$ is the
elastic energy of the cubic crystal in harmonic approximation.
In terms of homogeneous strains, $U_{TT}$ is given by Eq. (\ref{2.new28})
and $U_{QQT}$ by Eq. (\ref{4.23}). 
In principle a bilinear coupling term $U_{QT}$ 
between quadrupolar electronic and displacive degrees of 
freedom \cite{Kan}
should be included in expression (7.1).
This term which is known from Jahn-Teller phase transitions \cite{Geh}
can be essential if we want to describe transitions from a cubic
phase with quadrupolar disorder to non cubic phases with
ferro- quadrupolar order \cite{Lyn}.
However for the transition to the $Pa{\bar 3}$ phase, which
we identify with $\alpha$-Ce, the atomic center of mass positions
still occupy a face centered cubic lattice and the term $U_{QT}$
is irrelevant. Indeed, the driving force for the transition
$Fm{\bar 3}m \rightarrow Pa{\bar 3}$ is the quadrupole-quadrupole
interaction which becomes attractive (in reciprocal space)
at the $X$ point of the Brillouin zone. This fact leads to an
orientational order of quadrupolar electronic densities on four
different sublattices (Figs. 3, 4).
The term $U_{QQT}$ then prompts a lattice contraction at the first
order phase transition.
Notice that the term $U_{QT}$ is found to vanish for
a wave vector $\vec{q}$ at the $X$ point of the BZ.
We insist on these facts since within our view, the electronic
charge degrees of freedom, together with the lattice displacements,
are the driving forces of structural phase transitions in Ce
and related compounds.

The Hamiltonian (7.1) is not sufficient to describe the
magnetic phenomena that occur at the $\gamma-\alpha$
transition \cite{Mac,Kos}. In accordance with Kramers' theorem the
quadrupolar ordering, as described by Eq.~(7.1) is not
accompanied by a magnetic ordering.
We then conclude that the addition of an Anderson Hamiltonian
term $H_{cf}$ which takes into account the Friedel-Anderson
hybridization between conduction electrons and 4$f$ electrons
as well as the repulsive energy among the 4$f$ electrons
on a same site
is necessary (for a review see \cite{Ful}).
Such a Hamiltonian leads to the disappearance of local magnetic moments
below a characteristic temperature $T_K$.
The Kondo temperature $T_K$ increases with increasing 
hybridization matrix element $V$.
We then conclude that the
lattice contraction accompanying the $Pa{\bar 3}$
quadrupolar ordering 
or the quadrupolar order
enhances the hybridization $V$ and hence
increases $T_K$. This can lead to a situation where the
structural $\gamma \rightarrow \alpha$ transition and the
demagnetization of the 4$f$ state occur at a same temperature $T_1$.
Notice however that within this scenario the process is driven
by the structural (quadrupolar) transition at $T_1$ and
not by the Kondo volume collapse \cite{All2,Lav}.
In case where the enhancement of $T_K$ is insufficient, the
quadrupolar order and the concomitant
lattice contraction would occur at $T_1$ without
Kondo anomaly (disappearance of magnetic moment).
The condition $T_K < T_1$ does not ensure that the Kondo
anomaly actually occurs at lower $T$.
Quadrupolar ordering has been observed in a number of
Ce, Pr, Tm and U based compounds \cite{Mor1}.
A remarkable example is the magnetic semiconductor
TmTe with 4$f^{13}$ ($^2F_{7/2}$) electronic
configuration and N\'{e}el temperature $T_N$=0.43 K \cite{Las}. 
Although TmTe had been extensively studied before 1995,
the phenomenon of quadrupole ordering below $T_Q$=1.8 K was completely
overlooked \cite{Mat,Lin}. 
Numerous data on such compounds 
(CeAg \cite{Mor2}, CeB$_6$ \cite{Eff}, TmTe \cite{Mat}, 
TmAu$_2$ \cite{Kos2}, DyB$_2$C$_2$ \cite{Yam} etc.)
with quadrupole phase transitions show that
the magnetic ordering occurs at lower $T$.
A transition to an ordered magnetic phase in these
rare-earth intermetallic compounds indicates that the
Ruderman-Kittel-Kasuya-Yosida (RKKY) interaction
between localized magnetic moments prevails over
the the Kondo transition mechanism (for a review, see \cite{Bran}).
In zero magnetic field the magnetic susceptibility
shows no anomaly at $T_Q$ for TmTe \cite{Mat}, CeAg \cite{Mor2} 
and a very small anomaly for DyB$_2$C$_2$ \cite{Yam},
TmAu$_2$ \cite{Kos2}.
Within the present work we come to the conclusion that
quadrupolar order, as an electronic charge degrees of
freedom driven process on one hand, and magnetic properties 
(Kondo anomaly, magnetic order) on the other hand are
related indirectly via their coupling to lattice
displacements. 
An open question is a possible relation between quadrupolar order
and hybridization.
In our opinion a microscopic
derivation of the Anderson hybridization
Hamiltonian which should include details about
the symmetry of the lattice site and the electronic
orbitals constitutes a challenge for further work.

\acknowledgments 

We thank B. Lengeler, V.V. Moshchalkov, K. Parlinski, R.M. Pick and 
H. Wagner for
helpful criticism and discussions. We are grateful to
J.L. Sarrao and Z. Fisk for drawing our attention
to YbInCu$_4$.
A.V.N. thanks J.-M. Mignot and I.V. Solovyev for useful remarks.
This work has been financially supported by
the Fonds voor Wetenschappelijk Onderzoek, Vlaanderen.

\section {ERRATUM for Reference [21]} 
\label{sec:err} 

In Table 3 $-\beta$ (last column, third row)
should be replaced with $+\beta$.
In the last column of Table~5   $-\lambda$ (first row)
should be replaced with $+\lambda$, $-\mu$ (second row) - with $\nu$
and $-\nu$ (third row) - with $\mu$.
In Table 2 $i=7,8$ correspond to $\Gamma_7,2$ (third row, second column)
while $i=13,14$ - to $\Gamma_6$ (fifth row, second column).
In equation (4.7d) $c^{(2)}_3(hl)\, c^{(2)}_4(li)$ has to be replaced
by $c^{(2)}_2(il)\, c^{(2)}_2(li)$, on the right hand side 
of equation (5.11) in front
of summation it should stay $i$, not $\frac{i}{2}$.

\appendix
\section{} 
\label{sec:apA}

The Fourier transforms of the intersite electronic quadrupole-quadrupole
interactions are $3\times3$ matrices where the rows and
columns are labeled by the indices of the $T_{2g}$ functions
$S_2^k$, $k=1$, 2, 3.
The structure of these matrices depends on the symmetry of the
lattice and of the $T_{2g}$ functions.
The magnitude of the elements depends on the nature of the
electrons (localized 4$f$, or conduction 6$s$, 5$d$) and we
abbreviate the electronic indices $f$ or $s$, $d$ by the
label $A$, $B$ for $F=(ff)$ or $L=(l_1 l_2)$ writing
$v_k^A{}_{k'}^B$ for $v_k^{ff}{}_{k'}^{sd}$ etc..
We consider elements
\begin{eqnarray}
  v_k^A{}_{k'}^B(\vec{q})={\sum_{\vec{h \neq 0}}}' 
  v_k^A{}_{k'}^B(\vec{h})\, e^{i \vec{q} \cdot \vec{X}(\vec{h})} .
\label{A.1} 
\end{eqnarray}
Here $v_k^A{}_{k'}^B(\vec{h})$ refers to the elements
in real space with $\vec{h}=\vec{n}'-\vec{n}$.
Taking the site $\vec{n}$ as origin on a fcc lattice, the 
index $\vec{n}'$ (or $\vec{h}$) labels the twelve
neighbors.
Compare with expressions (\ref{2.6}), (\ref{3.9a}) and (\ref{3.16})
for the case $\Lambda=(T_{2g},k)$, $\Lambda'=(T_{2g},k')$.
Performing the lattice sums and using the symmetries of
the elements $v_k^A{}_{k'}^B(\vec{h})$ we obtain
\begin{eqnarray}
& &v^{A B}(\vec{q})= \label{A.2} \\
& &
{\scriptsize 
 {\hspace{-4mm}}4{\hspace{-1mm}}\left[ 
 \begin{array}{@{\hspace{-0.5mm}}c@{\hspace{-4mm}}c@{\hspace{-4mm}}c@{\hspace{-0.5mm}}}
 \gamma^{AB}\!C_{yz}\!+\!\alpha^{AB}\!(C_{zx}\!+\!C_{xy}\!) & -\beta^{AB}S_{xy} & -\beta^{AB}S_{zx} \\ 
 -\beta^{AB}S_{xy} & \gamma^{AB}\!C_{zx}\!+\!\alpha^{AB}\!(C_{xy}\!+\!C_{yz}\!) & -\beta^{AB}S_{yz} \\
 -\beta^{AB}S_{zx} & -\beta^{AB}S_{yz} & 
 \gamma^{AB}\!C_{xy}\!+\!\alpha^{AB}\!(C_{yz}\!+\!C_{zx}\!)    
\end{array} \right]
}
  \nonumber  
\end{eqnarray}
where $C_{ij}=\cos(q_i a/2) \cos(q_j a/2)$, and
$S_{ij}=\sin(q_i a/2) \\ \times \sin(q_j a/2)$.
Here $i,j$ stands for the Cartesian indices $x$, $y$, $z$, and $a$
is the cubic lattice constant.
Additional information on coupling matrices
can be found in Ref. \cite{Mic6} where a problem with similar symmetries
was considered for a molecular crystal.
The quantities $\gamma^{AB}$, $\alpha^{AB}$ and $\beta^{AB}$
stand for the matrix elements $v_k^A{}_{k'}^B(\vec{h})$
for $\vec{h}=a(1/2,1/2,0)$ with $(k=3,\,k'=3)$,
$(k=1,\,k'=1)$ and $(k=1,\,k'=2)$, respectively.
The interaction matrix $v^{AB}(\vec{q})$ has the largest
negative twofold degenerate eigenvalue at the $X$-point of the Brillouin zone.
For instance for $q_x^X=(2\pi/a)(1,0,0)$
$\lambda_{X_5^+}^{AB}=-4\gamma^{AB}$, where
$\gamma^{AB}=v_3^A{}_3^B(\vec{h})=v_2^A{}_2^B(\vec{h})>0$.
Hence the quadrupolar interaction parameters $\lambda_{X_5^+}^{AB}$ are
completely specified, their numerical
values are given in Table 2, fourth column.

\section{} 
\label{sec:apC}

We study the symmetry properties of the Bloch functions and 
rewrite Eq.~(\ref{3.1}) as
\begin{eqnarray}
 & &\langle \vec{R} | \, \vec{k},\alpha \rangle 
 =\psi_{\vec{k},\alpha}(\vec{R}) 
 \nonumber \\
 & &=
 \frac{1}{\sqrt{N}}
 \sum_{\vec{n}'} e^{i\vec{k} \cdot \vec{X}(\vec{n}')}
 \sum_{l \tau} \gamma_{l \tau}(\vec{k},\alpha) \,
 \phi_{l \tau}(\vec{R}-\vec{X}(\vec{n}')).
\label{C.1} 
\end{eqnarray}
Here $\phi_{l \tau}(\vec{r})={\cal R}_l(\vec{r}) S_l^{\tau}(\Omega)$
where $S_l^{\tau}(\Omega)$ are basis functions of the irreducible
representations of $O_h$. Here again $\tau \equiv (\Gamma, \mu, k)$.
We recall that
\begin{mathletters}
\begin{eqnarray}
 Y_l^m(\Omega)=\sum_{\tau} \beta_l^{m \tau} S_l^{\tau}(\Omega),
\label{C.2a} 
\end{eqnarray}
where $\beta_l$ is a unitary matrix. It follows that
\begin{eqnarray}
 \gamma_{l \tau}(\vec{k}, \alpha)=\sum_m \gamma_{lm}(\vec{k}, \alpha) 
 \beta_l^{m \tau}
\label{C.2b} 
\end{eqnarray}
\end{mathletters}
We observe that the expression (\ref{3.9b}) becomes
\begin{mathletters}
\begin{eqnarray}
  & &c_{k' l_1 l_2}(\vec{k},\alpha;\,\vec{p},\beta)
  \nonumber \\
  & &=\sum_{\tau_1\,\tau_2}
  \gamma_{l_1 \tau_1}^*(\vec{k},\alpha) \, 
  \gamma_{l_2 \tau_2}(\vec{p},\beta) \,
  c_{k'}(l_1 \tau_1,\,l_2 \tau_2), 
\label{C.3a} 
\end{eqnarray}
where
\begin{eqnarray}
 c_{k'}(l_1 \tau_1,\,l_2 \tau_2)=\int d\Omega\;
 S_{l_1}^{\tau_1}(\Omega)\, S_{k'}(\Omega) \,
 S_{l_2}^{\tau_2}(\Omega).
\label{C.3b} 
\end{eqnarray}
\end{mathletters}
The Bloch functions can be rewritten as basis functions of a
$d$-dimensional irreducible representation of the space group 
$Fm{\bar 3}m$ which is induced by a small representation 
$\Gamma_{\vec{k}_1}^{(p)}$ of the little group of a chosen
wave vector $\vec{k}_1$.
Here $p$ labels the small representation. One has $d=q \cdot t$,
where $q$ is the number of arms of the star of $\vec{k}_1$:
${}^* \! \vec{k}_1=\{ \vec{k}_1... \vec{k}_i... \vec{k}_q \}$,
and where $t$ is the dimension of the small representation.
(If $\vec{k}_1$ is not a high symmetry vector of the Brillouin zone,
$q=48$, the order of the group $O_h$, $t=1$ and $d=48$.)
The basis functions $\langle \vec{R}| \vec{k}_i,s_p,\alpha \rangle$,
where $s_p=1-t$ and $\vec{k}_i \in {}^* \! \vec{k}_1$,
all have the same energy. We can use the small representation
index $p$ to label the energies. Hence we will write
for the band index $\alpha$ the composite index $(p,\nu)$,
where $\nu$ labels the small representations $p$ that occur
more than once.
We consider (see also Appendix~B of Ref.~I)
\begin{eqnarray}
 \sum_{s_p=1}^t \sum_{i=1}^q 
 |\langle \vec{R} | \vec{k}_i, s_p, \alpha \rangle|^2 =
 {\frac {1}{N}} \sum_{\vec{n}'} 
 \phi^*_{l \tau}(\vec{r}_{\vec{n}'}) \phi_{l' \tau'}(\vec{r}_{\vec{n}'})
 \nonumber \\
 \times
 \sum_{s_p=1}^t \sum_{i=1}^q 
 \gamma^*_{l \tau}(\vec{k}_i,s_p,\alpha)\,
 \gamma_{l' \tau'}(\vec{k}_i,s_p,\alpha)
\label{C.4} 
\end{eqnarray}
for fixed $p$ and $\vec{k}_i \in {}^* \! \vec{k}_1$.
By the generalized Uns\"{o}ld theorem \cite{Tin} this expression
is an invariant of $Fm{\bar 3}m$ for any $\vec{r}_{\vec{n}'}$.
This means that the site symmetry is $O_h$ and the density
$\sum_k \phi^*_{l\, (\Gamma \mu k)}(\vec{r}_{\vec{n}'}) 
\phi_{l\, (\Gamma \mu k)}(\vec{r}_{\vec{n}'})$
stays invariant for any representation $\Gamma$ and index $\mu$
of $O_h$ (Uns\"{o}ld theorem).
We therefore obtain the condition
\begin{eqnarray}
 \sum_{s_p=1}^t \sum_{i=1}^q  
 \gamma_{l \tau}^*(\vec{k}_i,s_p,\alpha) \;
 \gamma_{l' \tau'}(\vec{k}_i,s_p,\alpha)
  \nonumber \\
   = \delta_{l l'}
  \delta_{\tau \tau'} \,
  f \! \left[ \vec{k}_1 \, \alpha;\, \Gamma,\mu
  \right],
\label{C.5} 
\end{eqnarray}
where $\vec{k}_1$ belongs to a basis domain $\Phi$ \cite{Bra}
and where
$f$ is a function of $\vec{k}_1$ and depends on the labels 
$\Gamma, \mu, \alpha$.

Now we calculate the single particle expectation values
$\langle (\rho_k^{l_1 l_2}(\vec{n}))^2 \rangle_0$, Eq.~(\ref{B.7}),
for conduction electrons.
From Eqs.~(\ref{3.11a},b) we have
\begin{eqnarray}
 & &\langle \rho_k^{l_1 l_2}(\vec{n}) 
 \rho_{k'}^{l'_1 l'_2}(\vec{n}) \rangle_0
  = {\frac {1}{N}} \sum_{\vec{q}} 
 \langle \rho_k^{l_1 l_2}(\vec{q})^{\dagger} 
 \rho_{k'}^{l'_1 l'_2}(\vec{q}) \rangle_0 
\nonumber \\ 
 & &= {\frac {1}{N^2}} \sum_{\vec{q}}  \sum_{\vec{p}} \sum_{\vec{h}}
 \langle a^{\dagger}_{\vec{p}-\vec{q}\, \alpha} a_{\vec{p} \beta} \,
 a^{\dagger}_{\vec{h}+\vec{q}\, \alpha'} a_{\vec{h} \beta'} \rangle_0
 \nonumber \\
 & & \times \, c^*_{k l_1 l_2}(\vec{p},\beta;\, \vec{p}-\vec{q}, \alpha) \,
 c_{k' l'_1 l'_2}(\vec{h}+\vec{q},\alpha';\, \vec{h}, \beta').
\label{B.1} 
\end{eqnarray}
Here summation is understood over the repeated band indices
$\alpha, \alpha',\beta, \beta'$.
The thermal expectation value is calculated with the single electron
Hamiltonian $U_0^c$, Eq. (\ref{A.6}).
We use standard Green's function (GF) techniques \cite{Zub}
and evaluate first the susceptibility (retarded GF) in the
static frequency limit ($z$=0)
\begin{eqnarray}
 & &\chi(z=0)=-\langle \! \langle
 a^{\dagger}_{\vec{p}-\vec{q}\, \alpha} a_{\vec{p} \beta}; \,
 a^{\dagger}_{\vec{h}+\vec{q}\, \alpha'} a_{\vec{h} \beta'}
 \rangle \! \rangle_{z=0} 
 \nonumber \\
 & &= -{\frac {n_{\vec{p}-\vec{q}\, \alpha}-n_{\vec{p}\, \beta}}
 {E(\vec{p}-\vec{q},\, \alpha)-E(\vec{p}, \beta)}}\;
 \delta_{\vec{p}-\vec{q}\; \vec{h}}\; 
 \delta_{\alpha\; \beta'}\; \delta_{\beta\; \alpha'}.
\label{B.3} 
\end{eqnarray}
Here $n_{\vec{p} \alpha}$ is the Fermi distribution function.
Therefrom we get the identity
\begin{eqnarray}
 n_{\vec{p}-\vec{q}\, \alpha}-n_{\vec{p}\, \beta}&=&
 (1-n_{\vec{p}\, \beta})\, n_{\vec{p}-\vec{q}\, \alpha} 
 \nonumber \\
 & & \times \, \left[1-e^{(E(\vec{p}-\vec{q},\, \alpha)-
 E(\vec{p}, \beta))/T} \right].
\label{B.4} 
\end{eqnarray}
We observe that in the high temperature case the static GF 
$\langle \! \langle A; B \rangle \! \rangle_{z=0}$ is related to
the thermal expectation value $\langle A\,B \rangle$ by
\begin{eqnarray}
 \langle A\,B \rangle_0=-T 
 \langle \! \langle A; B \rangle \! \rangle_{z=0}.
\label{B.5} 
\end{eqnarray}
Hence, for $|E(\vec{p}-\vec{q},\alpha)-E(\vec{p},\beta)|<T$, 
we obtain
\begin{eqnarray}
 & &\langle \rho_k^{l_1 l_2}(\vec{n}) \rho_{k'}^{l'_1 l'_2}(\vec{n}) \rangle_0
 = {\frac {1}{N^2}} \sum_{\alpha\, \beta} \sum_{\vec{p}\, \vec{q}}
 (1-n_{\vec{p}\, \beta})\, n_{\vec{p}-\vec{q}\, \alpha} 
 \nonumber \\
 &   & \times \,
 c^*_{k\, l_1 l_2}(\vec{p},\beta;\, \vec{p}-\vec{q}, \alpha) \,
 c_{k' l'_1 l'_2}(\vec{p},\beta;\, \vec{p}-\vec{q}, \alpha) .
\label{B.6} 
\end{eqnarray}
We next show that symmetry implies that $k=k'$ and $l_1=l'_1$,
$l_2=l'_2$.
Expression (\ref{B.6}) then reduces to (\ref{B.7}).
We rewrite expression (\ref{B.6}) as
\begin{eqnarray}
\langle \rho_k^{l_1 l_2}(\vec{n}) \rho_{k'}^{l'_1 l'_2}(\vec{n}) \rangle_0
  = {\frac {1}{N^2}} \sum_{\alpha\, \beta} \sum_{\vec{p}\, \vec{h}}
 (1-n_{\vec{p}\, \beta})\, n_{\vec{h}\, \alpha}\; 
 \nonumber \\ \times \,
 c^*_{k\, l_1 l_2}(\vec{p},\beta;\, \vec{h}, \alpha) \,
 c_{k' l'_1 l'_2}(\vec{p},\beta;\, \vec{h}, \alpha) .
\label{Bb.7} 
\end{eqnarray}
Using Eq. (\ref{C.5}), we now exploit the fact
that the summation over the Brillouin zone can be decomposed by
distinguishing the summation over the basic domain from the
summation over the components of irreducible representations.
Expression (\ref{Bb.7}) can be rewritten as
\begin{eqnarray}
\langle \rho_k^{l_1 l_2}(\vec{n}) \rho_{k'}^{l'_1 l'_2}(\vec{n}) \rangle_0
=\langle (\rho_k^{l_1 l_2}(\vec{n}) )^2 \rangle_0 \;
 \delta_{l_1\, {l'}_1}\, \delta_{l_2\, {l'}_2} \,
 \delta_{k\, k'},
\label{B.8} 
\end{eqnarray}
where
\begin{eqnarray}
 \langle (\rho_k^{l_1 l_2}(\vec{n}) )^2 \rangle_0=
 \sum_{\alpha\, \beta}  
 \sum_{\vec{h}_1 \in \Phi} 
 \sum_{\vec{p}_1 \in \Phi}
 \sum_{s_h} \sum_i
 \sum_{s_p} \sum_j
 \nonumber \\  
 \left|
 c_{k\, l_1 l_2}(\vec{h}_i, s_h, \alpha;\,  
 \vec{p}_j, s_p, \beta) 
 \right|^2 
 P(\vec{h}_1 \alpha;\, \vec{p}_1 \beta ) .
\label{B.9} 
\end{eqnarray}
Here we have taken into account the property (\ref{C.5}).
The quantity $P$ is a function of temperature due to the
presence of the Fermi distribution.
We next show that expression (\ref{B.9}) is independent of
the index of the quadrupolar component $k$
({\it i.e.} has the same value for $k$=1,2,3.)
We consider the summation over the components of a pair of
irreducible representations:
\begin{eqnarray}
 X &=& P(\vec{h}_1 \alpha;\, \vec{p}_1 \beta )\,
 \nonumber \\
 & \times & \sum_{s_h} \sum_i \sum_{s_p} \sum_j
 \left|
 c_{k\, l_1 l_2}(\vec{h}_i, s_h, \alpha;\,  
 \vec{p}_j, s_p, \beta)  \right|^2 .
\label{B.10} 
\end{eqnarray}
Inserting the explicit expression for $c_{k\, l_1 l_2}$ we obtain
\begin{eqnarray}
 X=\sum_{\Gamma_1\, \Gamma_2} \sum_{\mu_1\, \mu_2} 
 P'(\vec{h}_1, \alpha,\, \Gamma_1,\mu_1;\, \vec{p}_1, \beta,
 \, \Gamma_2,\mu_2)\, 
 \nonumber \\   \times \,
 \sum_{s_1 s_2} \left[
 c_k^*(l_1 \tau_1; l_2 \tau_2)  \right]^2,
\label{B.11} 
\end{eqnarray}
where $\tau_i=(\Gamma_i,\mu_i,s_i)$, $i=1,2$.
We have used relations (\ref{C.3a},b) and (\ref{C.5}).
The quantity $P'$ is defined by
\begin{eqnarray}
 & &P'(\vec{h}_1, \alpha,\, \Gamma_1,\mu_1;\, \vec{p}_1, \beta,
 \, \Gamma_2,\mu_2)
 \nonumber \\
 & &= P(\vec{h}_1 \alpha;\, \vec{p}_1 \beta )\;
 f \! \left[ \vec{h}_1 \, \alpha;\, \Gamma_1,\mu_1 \right] \,
 f \! \left[ \vec{p}_1 \, \beta;\, \Gamma_2,\mu_2 \right] .
\label{B.12} 
\end{eqnarray}
Finally we observe that as a consequence of the Wigner-Eckart
theorem
\begin{eqnarray}
 & & \sum_{s_1\, s_2}  
 c_k^*(l_1 \tau_1; l_2 \tau_2)\, c_{k'}(l_1 \tau_1; l_2 \tau_2)
 =\delta_{k\, k'} {\cal C}(\Gamma_1 \mu_1; \Gamma_2 \mu_2),
 \nonumber \\
 & & \label{B.12'} 
\end{eqnarray}
where ${\cal C}$ is independent of $k$
(as a square of a reduced matrix element).
This proves that $X$ and correspondingly
$\langle (\rho_k^{l_1 l_2}(\vec{n}) )^2 \rangle_0$
is independent of $k$.
Finally we quote a property of $\gamma_{l \tau}$
that follows from the completeness of the basis functions
\begin{eqnarray}
& & \sum_{\alpha} 
 \sum_{\vec{k}_1 \in \Phi} \sum_{s_p=1}^t \sum_{i=1}^q
 \gamma_{l \tau}^*(\vec{k}_i,s_p,\alpha)\,
 \gamma_{l' \tau'}(\vec{k}_i,s_p,\alpha)
  = \delta_{l\, l'}\, \delta_{\tau\, \tau'}\, N. 
  \nonumber \\
& & \label{B.14} 
\end{eqnarray}



\begin{references} 

\bibitem{Kos} 
D.C. Koskenmaki and K.A. Gschneidner, Jr., {\it Handbook on the 
Physics and Chemistry of Rare Earths}, ed. K. A. Gschneidner, Jr., 
and L. Eyring (Amsterdam: North-Holland, 1978), p. 337. 
 
\bibitem{Mal} 
D. Malterre, M. Grioni, Y. Baer, Adv. Phys. {\bf 45}, 299 (1996). 
 
\bibitem{Joh} 
B. Johansson, Phil. Mag. {\bf 30}, 469 (1974). 
  
\bibitem{Joh1} 
B.~Johansson, I.~A.~Abrikosov, M.~Alden, A.~V.~Ruban, H.~L.~Skriver,
Phys. Rev. Lett. {\bf 74}, 2335 (1995). 

\bibitem{Sva} 
A. Svane, Phys. Rev. B {\bf 53}, 4275 (1996).

\bibitem{Jar} 
T. Jarlborg, E. G. Moroni, and G. Grimvall, 
Phys. Rev. B {\bf 55}, 1288 (1997).
 
\bibitem{Laeg} 
J. L{\ae}gsgaard and A. Svane, Phys.~Rev. B {\bf 59}, 3450 (1999). 

\bibitem{All2} 
J. W. Allen, R. M. Martin, Phys. Rev. Lett. {\bf 49}, 1106 (1982); 
J. W. Allen, L. Z. Liu, Phys. Rev. B {\bf 46}, 5047 (1992). 

\bibitem{Lav} 
M. Lavagna, C. Lacroix, M. Cyrot, Phys. Lett. {\bf 90A}, 710
(1982); J. Phys. F {\bf 13}, 1007 (1985).
 
\bibitem{Gun} 
O. Gunnarsson, K. Sch\"{o}nhammer, Phys. Rev. Lett. {\bf 50}, 604 (1983); 
O. Gunnarsson, K. Sch\"{o}nhammer, Phys. Rev. B {\bf 28}, 4315 (1983). 
 
\bibitem{Bic} 
N. E. Bickers, D. L. Cox, J. W. Wilkins, Phys. Rev. B {\bf 36}, 2036 (1987). 

\bibitem{Mac} 
M.~R.~MacPherson, G.~E.~Everett, D.~Wohlleben, M.~B.~Maple,
Phys.~Rev.~Lett. {\bf 26}, 20 (1971).

\bibitem{And} 
P.~W.~Anderson, Phys.~Rev. {\bf 124}, 41 (1961). 

\bibitem{Joy} 
J. J. Joyce, A. J. Arko, J. Lawrence, P. C. Canfield, Z. Fisk, 
R. J. Bartlett, J. D. Thompson, Phys. Rev. Lett. {\bf 68}, 236 (1992).
  
\bibitem{Law} 
J. M. Lawrence,~A. J. Arko,~J. J. Joyce,~R. I. R. Blyth, R. J. Bartlett, 
P. C. Canfield, Z. Fisk, P. S. Riseborough,
Phys. Rev. B {\bf 47}, 15460 (1993).
 
\bibitem{Bly} 
R. I. R. Blyth, J. J. Joyce, A. J. Arko, P. C. Canfield, A. B. Andrews, 
Z. Fisk, J. D. Thompson, R. J. Bartlett,  P. Riseborough, J. Tang, 
J. M. Lawrence, Phys. Rev. B {\bf 48}, 9497 (1993); 
J. J. Joyce, A. B. Andrews, A. J. Arko, R. J. Bartlett, R. I. R. Blythe,
C. G. Olson, P. J. Benning, P. C. Canfield, D. M. Poirier,
Phys. Rev. B {\bf 54}, 17515 (1996).
 
\bibitem{Andr}
A. B. Andrews,~J. J. Joyce,~A. J. Arko,~J. D. Thompson,~J. Tang, 
J. M. Lawrence, J. C. Hemminger,
Phys. Rev. B {\bf 51}, 3277 (1995);
A. B. Andrews, J. J. Joyce, A. J. Arko, Z. Fisk,
ibid. {\bf 53}, 3317 (1996).

\bibitem{Ark}
A. J. Arko,~J. J. Joyce,~A. B. Andrews,~J. D. Thompson,
J. L. Smith, D. Mandrus, M. F. Hundley, A. L. Cornelius, E. Moshopoulou,
Z. Fisk, P. C. Canfield, A. Menovsky,
Phys. Rev. B {\bf 56}, R7041 (1997).

\bibitem{Cor}
A. L. Cornelius, J. M. Lawrence, J. L. Sarrao, Z. Fisk,
M. F. Hundley, G. H. Kwei, J. D. Thompson, C. H. Both,
F. Bridges, Phys. Rev. B {\bf 56}, 7993 (1997).

\bibitem{Sar}
J. L. Sarrao, Physica B {\bf 259-261}, 128 (1999), and references therein.

\bibitem{NM}
A. V. Nikolaev and K. H. Michel, Eur. Phys. J. B {\bf 9}, 619 (1999).

\bibitem{err1}
Expression (5.16) of Ref.~\cite{NM} should be replaced
with $\Lambda=-445$~K/a.u. ($-841$~K/{\AA}) which leads to 
$\triangle a=0.0022$ {\AA}. The corrected value is twice
the quantity $\triangle a$ reported in Ref.~\cite{NM}.

\bibitem{Mil}
A.~H.~Millhouse, A.~Furrer, Solid State Comm. {\bf 15}, 1303 (1974).

\bibitem{Mur1}
A.~P.~Murani, Z.~A.~Bowden, A.~D.~Taylor, R.~Osborn, W.~G.~Marshall,
Phys.~Rev.~B {\bf 48}, 13981 (1993).

\bibitem{Bra} 
 C.~J.~Bradley and A.~P.~Cracknell, 
 {\it The Mathematical Theory of Symmetry in Solids},
 (Clarendon, Oxford, 1972).

\bibitem{Cop}
J. R. D. Copley and K. H. Michel, J. Phys. Condens. Matter 
{\bf 5}, 4353 (1993).

\bibitem{Mic_LD}
K. H. Michel, D. Lamoen, W. I. F. David,
Acta Cryst. A{\bf 51}, 365 (1995).

\bibitem{Bet}
H. Bethe, Ann. Phys. (Germany) {\bf 3}, 133 (1929);
F. C. Von der Lage and H. A. Bethe, Phys. Rev. {\bf 71}, 612 (1947).

\bibitem{Zel}
P. Zielinski, K. Parlinski, J. Phys. C: Solid State Phys. {\bf 17},
3287 (1984).
\bibitem{Sco}
T.A. Scott, Phys. Rep. {\bf 27}, 89 (1976).
\bibitem{Dav}
W.I.F. David, R.M. Ibberson, T.J.S. Dennis, J.P. Hare,
and K. Prassides, Europhys. Lett. {\bf 18}, 219 (1992);
P.A. Heiney, G.B.M. Vaughan, J.E. Fischer, N. Coustel,
D.E. Cox, J.R.D. Copley, D.A. Neumann, W.A. Kamitakahara,
K.M. Creegan, D.M. Cox, J.P. McCauley, Jr.,
A.B. Smith III, Phys. Rev. B {\bf 45}, 4544 (1992).

\bibitem{Par}
K. Parlinski, private communication.
\bibitem{Mor1}
P. Morin and D. Schmitt, in {\it Ferromagnetic Materials},
K.H.J. Buschow and E.P. Wohlfarth Eds. (North-Holland, Amsterdam, 1990),
v. 5, p. 1.
\bibitem{Mor2}
P. Morin, J. Magn. Magn. Mater. {\bf 71}, 151 (1988).

\bibitem{Kan}
J. Kanamori, J. Appl. Phys. {\bf 31}, 145 (1960).
\bibitem{Geh}
G.A. Gehring and K.A. Gehring, Rep. Prog. Phys. {\bf 38}, 1 (1975).

\bibitem{Lyn}
R.M. Lynden-Bell and K.H. Michel, Rev. Mod. Phys. {\bf 66}, 721 (1994).

\bibitem{Ful}
P. Fulde, {\it Electron Correlations in Molecules and Solids},
Springer Heidelberg, 1995.

\bibitem{Las}
Y. Lassailly, C. Vettier, F. Holtzberg, A. Benoit, J. Flouquet,
Solid State Commun. {\bf 52}, 717 (1984).
\bibitem{Mat}
T. Matsumura, Y. Haga, Y. Nemoto, S. Nakamura, T. Goto, T. Suzuki,
Physica B {\bf 206-207}, 380 (1995);
T. Matsumura, S. Nakamura, T. Goto, H. Amitsuka, K. Matsuhira,
T. Sakakibara, T. Suzuki, J. Phys. Soc. Jpn {\bf 67}, 612 (1998).
\bibitem{Lin}
P. Link, A. Gukasov, J.-M. Mignot, T. Matsumura, T. Suzuki,
Phys. Rev. Lett. {\bf 80}, 4779 (1998).
\bibitem{Eff} 
J. M. Effantin, J.Rossat-Mignod, P. Burlet, 
H. Bartholin, S. Kunii, T. Kasuya, 
J. Magn. Magn. Mater. {\bf 47-48}, 145 (1985).
\bibitem{Kos2}
M. Kosaka, H. Onodera, K. Ohoyama, M. Ohashi, Y. Yamaguchi,
S. Nakamura, T. Goto, H. Kobayashi, S. Ikeda,
Phys. Rev. B {\bf 58}, 6339 (1998). 
\bibitem{Yam}
H. Yamauchi, H. Onodera, K. Ohoyama, T. Onimaru,
M. Kosaka, M. Ohashi, Y. Yamaguchi,
J. Phys. Soc. Jpn. {\bf 68}, 2057 (1999).
\bibitem{Bran}
N.B. Brandt and V.V. Moshchalkov, Adv. Phys. {\bf 33},
374 (1984).
%
\bibitem{Hir}
K. Hirota, N. Oumi, T. Matsumura, H. Nakao, Y. Wakabayashi,
Y. Murakami, Y. Endoh, Phys. Rev. Lett. {\bf 84}, 2706 (2000).

\bibitem{Mic6}  
 K.H. Michel, J.R.D. Copley, 
 Z. Phys. B Cond.~Matter {\bf 103}, 369 (1997).

\bibitem{Ash}
N. W. Ashcroft and N. D. Mermin, {\it Solid State Physics},
(Holt, Rinehart and Winston, 1976). 

\bibitem{Tin} 
 M.~Tinkham, 
 {\it Group Theory and Quantum Mechanics}, 
  (McGraw-Hill, New York, 1964). 

\bibitem{Zub} 
D. N. Zubarev, Usp. Fiz. Nauk {\bf 71}, 71 (1960)
[Sov. Phys. Usp. {\bf 3}, 320 (1960)].

\end{references}
\end{document}